\documentstyle[12pt]{article}
\newcommand{\mysection}[1]{\setcounter{equation}{0}\section{#1}} 
 
\def\be{\begin{equation}} 
\def\ee{\end{equation}} 
\def\l{\label} 
 
\def\F{{\cal F}} 
 
\def\S{{\cal S}} 
 
\def\T{{\cal T}} 
 
\def\Z{{}} 
\def\H{{\cal H}}

\def\W{{\cal W}}

\makeatletter\ifcase\@ptsize\font\teneufm=eufm10 
\font\seveneufm=eufm7\font\fiveeufm=eufm5 
\font\teneusm=eusm10\font\seveneusm=eusm7 
\font\fiveeusm=eusm5\or\font\teneufm=eufm10 scaled 
\magstephalf\font\seveneufm=eufm7\font\fiveeufm=eufm5 
\font\teneusm=eusm10 scaled\magstephalf 
\font\seveneusm=eusm7\font\fiveeusm=eusm5\or 
\font\teneufm=eufm10 scaled\magstep1\font\seveneufm=eufm7 
\font\fiveeufm=eufm5\font\teneusm=eusm10 scaled\magstep1 
\font\seveneusm=eusm7\font\fiveeusm=eusm5\fi 
 
\newfam\eufmfam\newfam\eusmfam\textfont\eufmfam=\teneufm 
\scriptfont\eufmfam=\seveneufm 
\scriptscriptfont\eufmfam=\fiveeufm 
\textfont\eusmfam=\teneusm\scriptfont\eusmfam=\seveneusm 
\scriptscriptfont\eusmfam=\fiveeusm 
 
\def\frak{\ifmmode\let\next\frak@\else 
\def\next{\errmessage{Use\string\frak\space only in math 
mode}}\fi\next}\def\frak@#1{{\frak@@{#1}}} 
\def\frak@@#1{\fam\eufmfam#1} 
\def\sh{\ifmmode\let\next\sh@\else 
\def\next{\errmessage{Use\string\sh\space only in math 
mode}}\fi\next}\def\sh@#1{{\sh@@{#1}}} 
\def\sh@@#1{\fam\eusmfam#1} 
 
\ifcase\@ptsize\font\tenmsa=msam10\font\sevenmsa=msam7 
\font\fivemsa=msam5\font\tenmsb=msbm10 
\font\sevenmsb=msbm7\font\fivemsb=msbm5\or 
\font\tenmsa=msam10 scaled\magstephalf 
\font\sevenmsa=msam7\font\fivemsa=msam5 
\font\tenmsb=msbm10 scaled\magstephalf 
\font\sevenmsb=msbm7\font\fivemsb=msbm5\or 
\font\tenmsa=msam10 scaled\magstep1\font\sevenmsa=msam7 
\font\fivemsa=msam5\font\tenmsb=msbm10 scaled\magstep1 
\font\sevenmsb=msbm7\font\fivemsb=msbm5\fi 
 
\newfam\msafam\newfam\msbfam\textfont\msafam=\tenmsa 
\scriptfont\msafam=\sevenmsa 
\scriptscriptfont\msafam=\fivemsa\textfont\msbfam=\tenmsb 
\scriptfont\msbfam=\sevenmsb 
\scriptscriptfont\msbfam=\fivemsb 
 
\def\Bbb{\ifmmode\let\next\Bbb@\else 
\def\next{\errmessage{Use\string\Bbb\space only in math 
mode}}\fi\next}\def\Bbb@#1{{\Bbb@@{#1}}} 
\def\Bbb@@#1{\fam\msbfam#1}\def\hexnumber@#1{\ifnum#1<10 
\number#1\else\ifnum#1=10 A\else\ifnum#1=11 
B\else\ifnum#1=12 C\else\ifnum#1=13 D\else\ifnum#1=14 E\else\ifnum#1=15 
F\fi\fi\fi\fi\fi\fi\fi} 
\def\msa@{\hexnumber@\msafam}\def\msb@{\hexnumber@\msbfam} 
\mathchardef\square="0\msa@03 
 
\makeatother 
\newcommand{\RR}{{\Bbb R}}

\newcommand{\SS}{{\Bbb S}} 
 
\begin{document} 
\begin{titlepage} 
 
\rightline{\tt hep-th/9909201} 
\rightline{MIT-CTP-2822} 
\rightline{UMN-TH-1741-99, TPI-MINN-99/6} 
\rightline{DFPD99/TH/10} 
 
\begin{center} 
 
{\large\bf Equivalence Principle, Higher Dimensional M\"obius Group and} 
 
\vspace{.333cm} 
 
{\large\bf the Hidden Antisymmetric Tensor of Quantum Mechanics} 
 
\vspace{.999cm} 
 
{\large Gaetano Bertoldi$^{1}$, Alon E. Faraggi$^{2}$ 
$\,$and$\,$ Marco Matone$^{3}$\\} 
\vspace{.2in} 
{\it $^{1}$ Center for Theoretical Physics\\ 
Laboratory for Nuclear Science and Department of Physics\\ 
Massachusetts Institute of Technology, Cambridge, MA 02139, USA\\ 
e-mail: bertoldi@ctp.mit.edu\\} 
\vspace{.08in} 
{\it $^{2}$ Department of Physics\\ 
University of Minnesota, Minneapolis MN 55455, USA\\ 
e-mail: faraggi@mnhepo.hep.umn.edu\\} 
\vspace{.08in} 
{\it $^{3}$ Department of Physics ``G. Galilei'' -- Istituto 
Nazionale di Fisica Nucleare\\ 
University of Padova, Via Marzolo, 8 -- 35131 Padova, Italy\\ 
e-mail: matone@pd.infn.it\\} 
 
\end{center} 
 
\vspace{.333cm} 
 
\centerline{\large Abstract} 
 
\vspace{.333cm}

\noindent 
We show that the recently formulated Equivalence Principle (EP) implies a 
basic cocycle condition both in Euclidean and Minkowski spaces, which holds in 
any dimension. This condition, that in one--dimension is sufficient to fix the 
Schwarzian equation \cite{6}, implies a fundamental higher dimensional 
M\"obius invariance which in turn univocally fixes the quantum version of the 
Hamilton--Jacobi equation. This holds also in the relativistic case, so that 
we obtain both the time--dependent Schr\"odinger equation and the 
Klein--Gordon equation in any dimension. We then show that the EP implies that 
masses are related by maps induced by the coordinate transformations 
connecting different physical systems. Furthermore, we show that the minimal 
coupling prescription, and therefore gauge invariance, arises quite naturally 
in implementing the EP. Finally, we show that there is an antisymmetric 
two--tensor which underlies Quantum Mechanics and sheds new light on the 
nature of the Quantum Hamilton--Jacobi equation. 
 
\noindent 
 
\vspace{.333cm} 
 
\noindent 
 
\end{titlepage}\newpage 
\setcounter{footnote}{0} 
\renewcommand{\thefootnote}{\arabic{footnote}} 
 
\mysection{Introduction}\l{intro} 
 
The consistent synthesis of the $20^{th}$ century most important philosophical
advances, Quantum Mechanics (QM) and General Relativity (GR), remains elusive. 
These two theories have changed the human experience of reality and allowed it 
to probe into the smallest and largest possible scales. Yet these two pillars 
of modern science remain incompatible at a fundamental level, despite enormous 
efforts devoted to formulating the proper mathematical theory that will 
embrace both QM and GR. It also seems that none of the current approaches to 
quantum gravity provides a satisfactory resolution. So, for example, the 
issues of the vacuum energy and generation of mass remain unsolved. Thus, it 
is fair to say that at present there does not exist a proper framework for the 
consistent formulation of quantum gravity, and what may be needed is a new 
paradigm. For example, one usually considers GR as the natural framework to 
describe gravitation seen as one of the four fundamental forces. On the other 
hand, QM is seen as the natural framework to describe interactions. So, the 
current view considers QM and GR as playing qualitatively rather different 
roles. 
 
Our view is going in another direction. Namely, suppose that QM and GR are in 
fact two facets of the same medal. If so, then we should need a reformulation 
of QM and a better understanding about the nature of GR and of the other 
interactions. Recently, in 
\cite{1}--\cite{6}, it has been proposed that QM can follow from an Equivalence 
Principle (EP) which is reminiscent of the Einstein EP. This principle 
requires that it is possible to connect all physical systems by coordinate 
transformations. In particular, there should always exist a coordinate 
transformation connecting a physical system with a non--trivial potential $V$ 
and energy $E$, to the one with $V-E=0$. Conversely, any allowed physical 
state should arise by a coordinate transformation from the state with $V-E=0$.
That is, under coordinate transformations, the trivial state should transform 
with an inhomogeneous term into a non--trivial one. In this context we stress 
that the EP has been formulated for states composed by one particle. However,
its formulation can be suitably generalized. 
 
The above aspects are intimately related with the concept of space--time. 
Actually, the removal of the peculiar degeneration arising in the classical 
concepts of rest frame and time parameterization is at the heart of the EP 
\cite{6}. In \cite{2,6} it was shown that this univocally leads to the Quantum 
Stationary HJ Equation (QSHJE). This is a third--order non--linear 
differential equation which provides a trajectory representation of QM. After 
publishing \cite{1}, the authors became aware that this equation was assumed 
in \cite{Floyd} as a starting point to formulate a trajectory interpretation 
of QM (see also \cite{RCarroll}). In \cite{4,6} it was shown that the 
trajectories depend on the Planck length through hidden variables which arise 
as initial conditions. So we see that QM may in fact need gravity. 
 
A property of the formulation is the manifest $p$--$q$ duality, which in turn 
is a consequence of the involutive nature of the Legendre transformation and 
of its recently observed relation with second--order linear differential 
equations \cite{7}. The role of the Legendre transformation in QM is related 
to the prepotential which appears in expressing the space coordinate in terms 
of the wave--function 
\cite{x}\cite{Carroll}\cite{DeAndreVancea}. 
 
The $p$--$q$ duality is deeply related to the M\"obius symmetry underlying the 
EP, which in turn fixes the QSHJE. This is also at the basis of energy 
quantization \cite{5,6}. In particular, the QSHJE is defined only if the ratio 
$w=\psi^D/\psi$ of a pair of real linearly independent solutions of the 
Schr\"odinger equation is a local homeomorphism of the extended real line 
$\hat\RR=\RR\cup\{\infty\}$ into 
itself. This is an important feature as the $L^2(\RR)$ condition, which in the 
Copenhagen formulation is a consequence of the axiomatic interpretation of the 
wave--function, directly follows as a basic theorem which only uses the 
geometrical glueing conditions of $w$ at $q=\pm\infty$ as implied by the EP. 
In particular, denoting by $q_-$ ($q_+$) the lowest (highest) $q$ for which 
$V(q)-E$ changes sign, we have that 
\cite{5,6} 
 
\vspace{.333cm} 
 
\noindent 
{\it If} 
\be 
V(q)-E\geq\left\{\begin{array}{ll}P_-^2 >0,&q<q_-,\\ P_+^2 >0,&q> 
q_+,\end{array}\right.
\l{perintroasintoticopiumeno}\ee 
{\it then $w=\psi^D/\psi$ is a local self--homeomorphism of $\hat\RR$ if and 
only if the corresponding Schr\"odinger equation has an 
$L^2(\RR)$ solution.} 
 
\vspace{.333cm} 
 
\noindent 
Thus, since the QSHJE is defined if and only if $w$ is a local 
self--homeomorphism of $\hat\RR$, this theorem implies that energy 
quantization {\it directly} follows from the QSHJE itself. Thus, we have that 
basic characteristics of QM are predicted by the EP as they arise by 
self--consistency from the EP without further assumptions. This is a 
fundamental aspect as in the standard formulation of QM the $L^2(\RR)$ 
condition is a consequence of the probabilistic interpretation of the 
wave--function. 
 
An important observation is that the Equivalence Postulate cannot be 
formulated consistently in Classical Mechanics (CM). To see this observe that 
if $\S_0^{cl}(q)$ and $\S_0^{cl\,v}(q^v)$ denote the classical Hamiltonian 
characteristic function, also called reduced actions, of two classical 
systems, then the coordinate transformation connecting the two systems can be 
defined by setting
\be
\S_0^{cl\,v}(q^v)=\S_0^{cl}(q),
\l{pcowcO}\ee
which implies
$\partial_{q^v}\S_0^{cl\,v}(q^v)=(\partial_{q^v}q)\partial_q\S_0^{cl}(q)$.
On the other hand, comparing the Classical Stationary Hamilton--Jacobi 
Equation (CSHJE) for $\S_0^{cl}(q)$ ($\W(q)\equiv V(q)-E$)
\be 
{1\over2m}\left({\partial\S^{cl}_0(q)\over\partial q} 
\right)^2+\W(q)=0, 
\l{CSHJEDabinitio}\ee 
with the CSHJE satisfied by $\S_0^{cl\,v}(q^v)$
\be 
{1\over2m}\left({\partial\S^{cl\,v}_0(q^v)\over\partial q^v} 
\right)^2+\W^v(q^v)=0, 
\l{CSHJEDabinitioB}\ee 
we see that $\W^v(q^v)=(\partial_{q^v}q)^2\W(q)$. This implies that the state 
corresponding to $\W=\W^0\equiv 0$ is a fixed a point, that is any coordinate 
transformation leaves $\W^0$ invariant as
$\W^0\to (\partial_{q^v}q)^2\W^0\equiv0$. This aspect can be also
understood by observing that in CM the state corresponding to 
$\W^0$ has a trivial reduced action, thus the transformation is highly singular 
in this case. This can be seen as the impossibility of implementing covariance 
of CM under the transformations defined by (\ref{pcowcO}). Thus, in CM it is 
not possible to generate all non--trivial states by a coordinate 
transformation from the trivial one. Consistent implementation of the EP 
requires a modification of CM. This univocally leads to QM. The starting point 
is to observe that the obstacle to the implementation of the EP is the 
transformation property $\W^v(q^v)=(\partial_{q^v}q)^2\W(q)$ which in turn is 
a consequence of the CSHJE. It follows that implementation of the EP has a 
highly dynamical content as it requires modifying the classical HJ equation. 
Therefore we should add to the CSHJE a still unknown term $Q$
\be 
{1\over2m}\left({\partial\S_0(q)\over\partial q}\right)^2+\W(q)+Q(q)=0, 
\l{aa10bbbxxxbprima}\ee 
where, in the $Q\to 0$ limit, $\S_0$ corresponds to the classical reduced 
action.

According to the EP, all physical systems composed by one particle under an 
external potential, labeled by the function 
$\W(q)\equiv V(q)-E$, can be connected by a coordinate transformation 
$q^a\rightarrow q^b=q^b(q^a)$, defined by
\be
\S_0^b(q^b)=\S_0^a(q^a).
\l{abinitio}\ee
Observe that at this stage we have not any dynamical information. It is just 
the implementation of the EP which will univocally fix the term $Q$ in 
(\ref{aa10bbbxxxbprima}). Furthermore, it is worth stressing that the 
equivalence concerns {\it all physical systems}. In particular, note that we 
are not restricting the equivalence to different energy levels of a system
with a fixed potential. The only restriction we are considering here concerns 
the number of particles composing the systems. As we said, we are considering 
the simplest case of systems composed by a single particle under an external 
potential. Nevertheless the EP can be suitably generalized to higher degrees 
of freedom.
 
It is immediate to see that the implementation of the EP has dramatic 
consequences. In fact, since the state with $\W=\W^0\equiv0$ corresponds to a 
fixed point, we see that the only way to implement the EP is to admit an 
inhomogeneous term in the transformation properties of $\W$ 
\be 
\W^a(q^a)\longrightarrow 
\W^b(q^b)=\left(\partial_{q^b}q^a\right)^2\W^a(q^a)+\Z(q^a;q^b). 
\l{ladoppiaww}\ee
On the other hand, by (\ref{aa10bbbxxxbprima}) and (\ref{abinitio}) we have 
$\W^b(q^b)+Q^b(q^b)=(\partial_{q^b}q^a)^2[\W^a(q^a)+Q^a(q^a)]$, so 
that
\be 
Q^a(q^a)\longrightarrow 
Q^b(q^b)=\left(\partial_{q^b}q^a\right)^2Q^a(q^a)-\Z(q^a;q^b).
\l{laQQQ9}\ee 
We used the notation $\Z(q^a;q^b)$ to stress that the unknown term depends on 
the functional relation between $q^a$ and $q^b$. The fundamental fact is that 
this term is fixed by the basic cocycle condition 
\be 
\Z(q^a;q^c)=\left(\partial_{q^c}q^b\right)^2\left[\Z(q^a;q^b)-\Z(q^c;q^b) 
\right], 
\l{inhomtrans}\ee 
which follows as consistency condition for (\ref{ladoppiaww}) or, 
equivalently, (\ref{laQQQ9}). Actually, in one dimension a key point was the 
following result \cite{6}
 
\vspace{.333cm} 
 
\noindent 
{\it The cocycle condition (\ref{inhomtrans}) uniquely defines the Schwarzian 
derivative up to a multiplicative constant and a coboundary term.} 
 
\vspace{.333cm} 
 
\noindent 
In particular, one obtains $\Z(q^a;q^b)=-\beta^2\{q^a,q^b\}/4m$, where 
$\{f,q\}=f'''/f'-3(f''/f')^2/2$ denotes the Schwarzian derivative 
and $\beta$ is a constant with the dimension of an action. Since in the 
classical case the term $\Z(q^a;q^b)$ must disappear from (\ref{ladoppiaww}), 
we have that the classical limit is reached for $\beta\to0$. Thus $\beta$ is 
naturally identified with 
$\hbar$. Furthermore, one sees that 
the inhomogeneous term $\Z(q^a;q^b)$ has a purely quantum origin\footnote{We 
refer to \cite{6} for several explicit examples of the formulation.}. 

An important issue of the present formulation concerns the similarity between 
the postulate equivalence of states and the Einstein EP. According to the 
Einstein EP it is always possible to choose a locally inertial coordinate 
system such that the physical laws have the same form as in unaccelerated 
coordinate systems in the absence of gravitation. The EP we formulated states 
that it is always possible to choose a a coordinate system in such a way that 
the reduced action corresponds to the one of the free particle with vanishing 
energy. While in the case of the Einstein EP, it is the {\it gravitational 
field} which is ``locally balanced" by a coordinate transformation, here there 
is an {\it arbitrary external potential} which is ``globally balanced" by a 
coordinate transformation. Another fundamental difference concerns the 
framework in which this is formulated. While the Einstein EP is formulated at 
the level of the equation of motions, here the formulation is implemented in 
the framework of HJ theory. This is a quite crucial difference. This becomes 
particularly transparent if we consider the case of a time--independent 
potential. In this case we can use the reduced action so that time never 
appears directly in the relevant equations. Only after the QSHJE is solved, 
one introduces time parameterization according to Jacobi theorem \cite{Floyd}, 
that is $t-t_0=\partial_q\S_0$. The fact that the QSHJE differs from the 
classical version implies that the conjugate momentum 
$\partial_q\S_0$ does not coincide with the mechanical one $m\dot q$. Thus, a 
feature of the EP is that time arises as parameter for trajectories and it is 
not introduced a priori. We believe this is a distinctive feature of HJ theory 
whose power fully manifests in the present formulation of QM. The different 
role of time in the formulation of the two equivalence principles can be also 
seen in considering the equation of motion of a particle in an external 
gravitational field $m\ddot q=mg$. Performing the time--dependent coordinate 
transformation $q'=q-gt^2/2$, we have $m\ddot q'=0$, for any value of the 
energy $E$ of the particle, including the free particle at rest for which 
$E=0$. So that, depending on the initial conditions of
$m\ddot q'=0$, we may have $q'$ to be constant, say $q'=0$. Therefore, there
are no selected frames if one uses time--dependent coordinate transformations. 
In other words, while the classical reduced action, which is not a function of 
time, is trivial, the equation of motions contain the time parameter which 
continues to flow. Hence, while with the CSHJE description it is not always 
possible to connect two systems by a coordinate transformation, this is not 
the case if one describes the dynamics using Newton's equation. In particular, 
in finding the coordinate transformation reducing the CSHJE description of 
$m\ddot q=mg$, one has
\be
{1\over2m}\left({\partial\S_0^{cl}(q)\over\partial q}\right)^2-mgq+E=0,
\l{inboccaallupoChiara}\ee
for which there is no coordinate transformation $q\longrightarrow q^v(q)$ such 
that $\S_0^{cl\,v}(q^v)=\S_0^{cl}(q)$ with $\S_0^{cl\,v}(q^v)$ the reduced 
action of the free particle with $E=0$. Time parameterization can be seen as a 
way to express a constant, say $0$, by means of the solution of the equation 
of motions, $q=f(t)$. For example, for a particle with constant velocity, we 
have $0=q-vt$, so that particle's position can be denoted by either 
$q$ itself or $vt$. In this way one can always reduce to the particle at rest 
by simply setting $q'=q -f(t)$. While in the case of the CSHJE description 
there is the degenerate case $cnst=mvq$, corresponding to 
$\S_0^{cl\,v}(q^v)=\S_0^{cl}(q)$, time parameterization provides a 
well--defined and invertible transformation {\it i.e.} 
$q'=q-f(t)\longrightarrow q=q'+f(t)$. The reason underlying the differences in 
considering the role of space and time is that fixed values of
$q$ and $t$ correspond to quite different situations. Even if the particle is
at rest, say at $q=0$, {\it time continues to flow}. It is just the use of time
that allows to connect different systems by a coordinate transformation.

There is a common feature underlying both the Eistein EP and the one we 
formulated. Namely, note that the existence of the classical systems with 
$\S_0^{cl}=0$, is essentially the reason of the impossibility of implementing
covariance of CM under the transformations defined by (\ref{pcowcO}). To be 
more precise, note that Eqs.(\ref{CSHJEDabinitio}) and (\ref{CSHJEDabinitioB}) 
can be seen as a first step in checking covariance. However, these equations 
have not any particular content. The problem of covariance arises when one 
tries to connect them by some transformation. We have seen that there is an 
inconsistency if we consider the coordinate transformation 
$\S_0^{cl\,v}(q^v)=\S_0^{cl}(q)$. It is just the removal of this inconsistency
which allows the implementation of the EP and therefore to have covariance. 
This univocally leads to QM. Thus, similarly to GR which can be derived by 
implementing the principle of general covariance under diffeomorphisms, also 
QM arises from a covariance principle. In our case covariance essentially 
relies on the request that the transformation 
$\S_0^v(q^v)=\S_0(q)$ be always defined. This immediately discards CM and 
modifies the classical concept of particle at rest. Thus we see that, 
similarly to the case of GR, the EP implies a principle of covariance.

One may wonder whether the properties of the Schwarzian derivative, an 
intrinsically one--dimensional (possibly complex) object, extend to higher 
dimension. Experience with string theory and CFT would indicate that similar 
properties are in fact strictly related to low--dimensional spaces. 
Nevertheless, these are related to the appearance of the QSHJE, and so, since 
the essence of QM manifests itself already in one--dimension, one may in fact 
believe that the higher dimensional generalization exists. We will in fact 
show that the basic fact underlying the construction is that the EP implies a 
M\"obius symmetry in any dimension. More precisely, the EP implies the higher 
dimensional analogue of the cocycle condition. 
 
One of the main results of the present paper is the proof that the above 
condition leads, in the case of the Euclidean metric, to an invariance under 
D--dimensional M\"obius transformations. In the case of the Minkowski metric 
the relevant invariance is with respect to the (D+1)--dimensional conformal 
group. This result is also non--trivial from
 the mathematical point of view, and may have implications for the higher 
dimensional diffeomorphisms. This M\"obius symmetry will then univocally lead 
to the time--dependent Schr\"odinger equation in higher dimension. 
 
Remarkably, we will see that the EP in fact implies also the higher dimensional
Relativistic Quantum HJ Equation (RQHJE) with external potentials. 
Furthermore, while considering an external potential leads to a mixing between 
the kinetic and potential part in deriving the RQHJE, this equation is 
obtained quite naturally once one considers the minimal coupling prescription. 
This aspect is a relevant feature of the EP which in fact corresponds to a 
sort of naturalness. Namely, the right framework to formulate it is the exact 
one, that is special relativity. So, for example, the time--dependent 
Schr\"odinger equation is simply derived as the non--relativistic 
approximation from the RQHJE. Furthermore, as we will see in the case of the 
Klein--Gordon equation in the presence of the electromagnetic field, the 
minimal coupling prescription is in fact the natural one. This indicates that 
gauge theories are deeply related to the EP. In this context, it is useful to 
stress that the standard Schr\"odinger problems one usually considers 
correspond to ideal situations. So for example, a potential well cannot be 
seen as a fundamental interaction. Actually, the Schr\"odinger problems one 
may consider at the level of fundamental interactions essentially concern the 
electromagnetic one. It is then interesting that the Schr\"odinger equation 
for minimal coupled potentials simply follows from the EP as a 
non--relativistic limit. 
 
Another key ingredient in the one dimensional derivation of QM from the EP was 
the following identity involving the Schwarzian derivatives 
\be 
(\partial_q\S_0)^2={\hbar^2\over2}(\{e^{{2i\over\hbar}\S_0},q\}-\{\S_0,q\}). 
\nonumber\ee 
Again, in the present paper we will find the generalization of this identity 
to higher dimension and in the relativistic case. 
 
We started the introduction by arguing for the need for a radical new paradigm 
for QM. The fact that QM arises from the EP may suggest that masses have a 
quantum origin. We will show that indeed this may be the case. The point is 
that in the relativistic case one has 
\be 
\W={1\over 2}mc^2, 
\l{pdoj2}\ee 
then we have that mass of a particle is obtained from the state corresponding to 
$\W^0\equiv0$ and is due to the inhomogeneous term which arises from coordinate 
transformations. 
 
Another basic feature of the present approach concerns the appearance of a new 
field which underlies QM. This is one of the new points one meets in 
considering the higher dimensional generalization of our formulation. As we 
will see, this field arises by solving the continuity equation associated to 
the QHJE. In particular, this equation defines a 
$(D-2)$--form which in turn defines an antisymmetric 2--tensor. 
 
Our paper is organized as follows. In section \ref{lasezione2} we set the 
notation and derive the higher dimensional cocycle condition. In section 
\ref{mobius} we will prove the invariance of the cocycle condition under the 
D--dimensional M\"obius transformations. In section \ref{lascr} we derive the 
higher--dimensional Schr\"odinger equation and discuss a possible connection 
with the holographic principle. In section 
\ref{questasezione} we then discuss the generalization to the relativistic case.
We show that in the case of the Minkowski metric, the cocycle condition is 
invariant under the (D+1)--dimensional conformal group. We derive the 
generalization of our approach for the Klein--Gordon equation and show how the 
time--dependent non--relativistic limit correctly reproduces the 
time--dependent Schr\"odinger equation. In section \ref{gauge} we discuss the 
generalization to the case with a four--vector including covariant 
derivatives. We also investigate the generation of mass in our approach and 
the appearance of the hidden antisymmetric tensor field which underlies QM.  
Finally, Appendices A and B are devoted to some more technical aspects of 
sections \ref{mobius} and 
\ref{questasezione}.
 
\mysection{EP and cocycle condition}\l{lasezione2} 
 
Let us consider the case of two physical systems with Hamilton's 
characteristic functions $\S_0$ and $\S_0^v$ and denote the coordinates of the 
two systems by $q$ and $q^v$ respectively. Let us set 
\be 
\S_0^v(q^v)=\S_0(q). 
\l{is1}\ee 
Observe that there are no particular assumptions in making the above 
identification. The point is that the physical content is in the functional 
dependence of 
$\S_0^v$ and $\S_0$ on their arguments $q^v$ and $q$ respectively, and (\ref{is1})
simply defines a functional relation between $q^v$ and $q$. One may also 
choose another rule connecting $\S_0^v(q^v)$ and $\S_0(q)$. However, as the 
one--dimensional case shows  
\cite{1}--\cite{6}, the formulation would result much more cumbersome. Thus, in a  
certain sense, we can say that $\S_0$ transforms as a scalar. In particular, 
the true assumption which underlies (\ref{is1}) is that there is a functional 
relation between  the coordinates of two arbitrary physical systems 
characterized by the systems  themselves. This is essentially the content of 
the EP we will formulate. Note that the  existence of a non--singular 
functional relation between the coordinate of two physical  systems cannot 
hold for all states of CM. In fact, (\ref{is1}) does not make sense once  one 
considers the classical state with  
$\W=0$, corresponding to  
$\S_0(q)=cnst$. Thus, requiring that (\ref{is1}) is defined for all systems 
implies that $\S_0(q)=cnst$ cannot corresponds to a physical state. This 
corresponds to a criticism of the concept of rest frame in CM. It is just the 
removal of the peculiar  degeneration arising in the classical concepts of 
rest frame and time parameterization,  discussed in great detail in sect.2 of 
Ref.\cite{6}, which provides the physical  motivation for formulating the EP. 
 
As in the one--dimensional case, we will see that the implementation of the 
EP, not only excludes CM, but also uniquely leads to the quantum version of 
the HJ equation. 
 
Note that Eq.(\ref{is1}) induces, in the one--dimensional case, the map  
\be 
q\longrightarrow q^v=v(q), 
\l{6xxx}\ee 
where 
\be 
v=\S_0^{{v}^{\;-1}}\circ\S_0, 
\l{10s}\ee 
with $\S_0^{{v}^{\;-1}}$ denoting the inverse of $\S_0^v$. This construction 
is equivalent to say that the map (\ref{6xxx}) induces the transformation 
$\S_0\longrightarrow\S_0^v=\S_0\circ v^{-1}$, 
that is $\S_0(q)\longrightarrow\S_0^v (q^v )=\S_0(q( q^v ))$. 
 
In the higher dimensional case, the relation $\S_0^v(q^v)=\S_0(q)$ defines 
infinitely many maps $q\longrightarrow q^v=v(q)$.\footnote{Tensorial 
properties are characterized  by giving specific rules under given 
transformations of coordinates. Here we are not  giving the transformation 
rules of $\S_0$ under a set of coordinate transformations.  Rather, we are 
defining a set of coordinate transformations starting from the knowledge  of 
the functional structure of $\S_0^v(q^v)$ and $\S_0(q)$. For this reason, 
strictly  speaking, even if $\S_0^v(q^v)=\S_0(q)$, the reduced action cannot 
be considered a  scalar function.}  Since, as we will see, the EP requires 
that two arbitrary physical  systems can be always connected by a coordinate 
transformation, the only condition we  need is that there exists the inverse 
of the map $v$. This is not sufficient to fix the particular form of the 
$v(q)$. However, as we will see, we only need that for any pair of states 
there exists a invertible map  (\ref{6xxx}) satisfying (\ref{is1}). We will 
call such maps  
$v$--transformations. 
 
One of the main results in \cite{1}--\cite{6} was that the reduced action 
$\S_0^0(q^0)$ corresponding to the free system with vanishing 
energy\footnote{We note that a common shift of $V$ and $E$ by a constant does 
not change $\W\equiv V-E$. Since this is the combination in which the data $V$
and $E$ enter in the equation of motions, we see that the case 
$V-E=0$ is indistinguishable from $V=E=0$.} is not a constant but the
``self--dual state'' 
\be 
e^{{2i\over\hbar}\S_0^0}={q^0+i\bar\ell_0\over q^0-i\ell_0}, 
\l{ints000}\ee 
with $\ell_0$, ${\rm Re}\,\ell_0\ne0$, a complex integration constant. This 
corresponds to the overlooked zero mode of the conformal factor in the quantum 
analogue of the Hamilton--Jacobi equation \cite{1}\cite{2}\cite{6}. 
Furthermore, in Ref.\cite{1} the function $\T_0(p)$, defined as the Legendre 
transform of the reduced action, was introduced 
\be 
\T_0(p)=\sum_{k=1}^Dq_kp_k-\S_0(q),\qquad\S_0(q)=\sum_{k=1}^Dp_kq_k-\T_0(p). 
\l{y1}\ee 
While $\S_0(q)$ is the momentum generating function, its Legendre dual 
$\T_0(p)$ is the coordinate generating function 
\be 
p_k={\partial\S_0\over\partial q_k},\qquad q_k={\partial\T_0\over\partial p_k}. 
\l{y2}\ee 
 
Let us now consider the Classical Stationary Hamilton--Jacobi Equation (CSHJE) 
in $D$--dimensions 
\be 
{1\over2m}\sum_{k=1}^D\left({\partial\S^{cl}_0\over\partial q_k} 
\right)^2+\W(q)=0, 
\l{CSHJED}\ee 
where 
\be 
\W(q)\equiv V(q)-E, 
\l{zummoloa1}\ee 
with $V(q)$ the potential and $E$ the energy. We denote by $\H$ the space of 
all possible $\W$'s corresponding to physical systems composed by one particle
(the extension to more general cases will be investigated elsewhere). 
 
In \cite{1} the following Equivalence Principle has been formulated 
 
\vspace{.333cm} 
 
\noindent 
{\it For each pair $\W^a,\W^b\in\H$, there is a $v$--transformation such that} 
\be 
\W^a(q)\longrightarrow{\W^a}^v(q^v)=\W^b(q^v). 
\l{equivalence}\ee 
We will see that the implementation of the EP will univocally lead to the 
QSHJE. 
 This implies that there always exists the trivializing coordinate $q^0$ 
for which 
$\W(q)\longrightarrow\W^0(q^0)$, where 
\be 
\W^0(q^0)=0. 
\l{dwpo}\ee 
In particular, since the inverse transformation should exist as well, it is 
clear that the trivializing transformation should be locally invertible. We 
will also see that since classically $\W^0$ is a fixed point, implementation 
of (\ref{equivalence}) requires that 
$\W^b(q^v)$ is given in terms of $\W^a(q)$ (times a suitable Jacobian) together 
with an additive term. In other words, the EP immediately implies that the 
$\W$ states transform inhomogeneously. 
 
The fact that the EP cannot be consistently implemented in Classical Mechanics 
(CM) is true in any dimension. To show this let us consider the coordinate
transformation induced by the identification
\be
\S_0^{cl\, v}(q^v)=\S_0^{cl}(q).
\l{pcowcOII}\ee
Then note that the CSHJE 
\be 
{1\over2m}\sum_{k=1}^D({\partial_{q_k}\S_0^{cl}(q)})^2+\W(q)=0, 
\l{012}\ee 
provides a correspondence between $\W$ and $\S_0^{cl}$ that we can use to fix, 
by consistency, the transformation properties of $\W$ induced by that of $\S_0 
$. In particular, since $\S_0^{cl\, v}(q^v)$ must satisfy the 
CSHJE 
\be 
{1\over2m}\sum_{k=1}^D({\partial_{q^v_k}^v\S_0^{cl\,v}(q^v)})^2+\W^v(q^v)=0, 
\l{zummoloa30}\ee 
by (\ref{pcowcOII}) we have
\be 
p_k\longrightarrow p^v_k={\partial\S_0^{cl\,v}(q^v)\over\partial 
q^v_k}=\sum_{i=1}^D{\partial q_i\over\partial 
q^v_k}{\partial\S_0^{cl}(q)\over\partial q_i}= 
\sum_{i=1}^DJ_{ki}p_i, 
\l{insomma}\ee 
where $J$ is the Jacobian matrix 
\be 
J_{ki}={\partial q_i\over\partial q^v_k}. 
\l{zummoloa12}\ee 
Let us introduce the notation 
\be 
(p^v|p)={\sum_k p_k^{v^2}\over\sum_kp_k^2}={p^tJ^tJp\over p^tp}. 
\l{natura2}\ee 
Note that in the $1$--dimensional case 
\be 
(p^v|p)=\left({p^v\over p}\right)^2=\left({\partial\S_0\over\partial q^v} 
{\partial q\over\partial\S_0}\right)^2=\left({\partial q^v\over\partial q} 
\right)^{-2}, 
\l{zummoloa13}\ee 
so that the ratio of momenta corresponds to the Jacobian of a coordinate 
transformation. By (\ref{012}), we have 
\be 
\W(q)\longrightarrow\W^v(q^v)=(p^v|p)\W(q), 
\l{piccolo}\ee 
that for the $\W^0$ state gives 
\be 
\W^0(q^0)\longrightarrow\W^v(q^v)=(p^v|p^0)\W^0(q^0)=0. 
\l{qddd}\ee 
Thus we have \cite{1} 
 
\vspace{.333cm} 
 
\noindent 
{\it $\W$ states transform as quadratic differentials under classical 
$v$--maps. It follows that $\W^0$ is a fixed point in $\H$. Equivalently, in 
CM the space 
$\H$ cannot be reduced to a point upon factorization by the 
classical $v$--transformations. Hence, the EP (\ref{equivalence}) cannot be 
consistently implemented in CM. This can be seen as the impossibility of 
implementing covariance of CM under the coordinate transformation defined by 
(\ref{pcowcOII}).} 
 
\vspace{.333cm} 
 
It is therefore clear that in order to implement the EP we have to deform the 
CSHJE. As we will see, this requirement will determine the equation for $\S_0$ 
in any dimension. 
 
Let us discuss its general form. First of all observe that adding a constant to 
$\S_0$ does not change the dynamics. Actually, Eqs.(\ref{y1})(\ref{y2}) are 
unchanged upon adding a constant to either $\S_0$ or $\T_0$. Then, the most 
general differential equation $\S_0$ should satisfy has the structure 
\be 
\F(\nabla\S_0,\Delta\S_0,\ldots)=0. 
\l{traslazione}\ee 
\noindent 
Let us write down Eq.(\ref{traslazione}) in the general form 
\be 
{1\over2m}\sum_{k=1}^D({\partial_k\S_0(q)})^2+\W(q)+Q(q)=0. 
\l{aa10bbbb}\ee 
The transformation properties of $\W+Q$ under the $v$--maps (\ref{6xxx}) are 
determined by the transformed equation 
\be 
{1\over2m}\sum_{k=1}^D({\partial_k\S_0^v(q^v)})^2/2m+\W^v(q^v)+Q^v(q^v)=0, 
\l{labella998}\ee 
which by (\ref{is1}) and (\ref{aa10bbbb}) yields 
\be 
\W^v(q^v)+Q^v(q^v)=(p^v|p)\left[\W(q)+Q(q)\right]. 
\l{yyyxxaa10bbbb}\ee 
 
A basic guidance in deriving the differential equation for $\S_0$ is that in 
some limit it should reduce to the CSHJE. In \cite{1}\cite{2} it was shown 
that the parameter which selects the classical phase is the Planck constant. 
Therefore, in determining the structure of the $Q$ term we have to take into 
account that in the classical limit 
\be 
\lim_{\hbar\to0}Q=0. 
\l{classicoqezero}\ee 
 
The only possibility to reach any other state $\W^v\ne0$ starting from $\W^0$ 
is that it transforms with an inhomogeneous term. Namely as 
$\W^0\longrightarrow 
\W^v(q^v)\ne0$, it follows that for an arbitrary $\W^a$ state 
\be 
\W^v(q^v)=(p^v|p^a)\W^a(q^a)+\Z(q^a;q^v), 
\l{azzoyyyxxaa10bbbb}\ee 
and by (\ref{yyyxxaa10bbbb}) 
\be 
Q^v(q^v)=(p^v|p^a)Q^a(q^a)-\Z(q^a;q^v). 
\l{azzo2yyyxxaa10bbbb}\ee 
Let us stress that the purely quantum origin of the inhomogeneous term
$\Z(q^a;q^v)$ is particularly transparent once one consider the 
compatibility between the classical limit (\ref{classicoqezero}) and the 
transformation properties of $Q$ in Eq.(\ref{azzo2yyyxxaa10bbbb}). 
 
The $\W^0$ state plays a special role. Actually, setting $\W^a=\W^0$ in 
Eq.(\ref{azzoyyyxxaa10bbbb}) yields 
\be 
\W^v(q^v)=\Z(q^0;q^v), 
\l{ddazzoyyyxxaa10bbbb}\ee 
so that, according to the EP (\ref{equivalence}), all the states correspond to 
the inhomogeneous part in the transformation of the $\W^0$ state induced by 
some $v$--map. 
 
Let us denote by $a,b,c,\ldots$ different $v$--transformations. Comparing 
\be 
\W^b(q^b)=(p^b|p^a)\W^a(q^a)+\Z(q^a;q^b)=\Z(q^0;q^b), 
\l{ganzate}\ee 
with the same formula with $q^a$ and $q^b$ interchanged we have 
\be 
\Z(q^b;q^a)=-(p^a|p^b)\Z(q^a;q^b), 
\l{inparticolare}\ee 
in particular 
\be 
\Z(q;q)=0. 
\l{inparticolareb}\ee 
More generally, comparing 
\be 
\W^b(q^b)=(p^b|p^c)\W^c(q^c)+\Z(q^c;q^b)=(p^b|p^a)\W^a(q^a)+ 
(p^b|p^c)\Z(q^a;q^c)+\Z(q^c;q^b), 
\l{zummoloa14}\ee 
with (\ref{ganzate}) we obtain the basic cocycle condition 
\be 
\Z(q^a;q^c)=(p^c|p^b)\left[\Z(q^a;q^b)-\Z(q^c;q^b)\right], 
\l{cociclo3}\ee 
which expresses the essence of the EP. 
 
\mysection{$\Z(q^a;q^b)$ and the higher dimensional M\"obius group}\l{mobius}
 
In this section, we will show that $\Z(q^a;q^b)$ vanishes identically if $q^a$ 
and $q^b$ are related by a M\"obius transformation. This is a consequence of 
Eqs.(\ref{traslazione}) and (\ref{cociclo3}) and generalizes the 
one--dimensional result obtained in \cite{1}\cite{2}\cite{6}, which states 
that 
$\Z(q^b;q^a)=0$ if and only if $q^b$ is a linear fractional transformation of 
$q^a$ 
\be 
q^b={Aq^a+B\over Cq^a+D}\, ,\quad\qquad AD-BC\not=0. 
\l{bosse1}\ee 
As in the one dimensional case, the M\"obius symmetry will fix the $Q$--term 
in Eq.(\ref{aa10bbbb}). Before going into the details of the proof, we will 
give a brief overview of the M\"obius group (see, for example, 
\cite{Vuorinen}). 
 
\subsection{Higher dimensional M\"obius group} 
 
Let us denote by $q=(q_1,\cdots,q_D)$ an arbitrary point in $\RR^D$. A {\it 
similarity} is the affine mapping 
\be 
q\longrightarrow Mq+b, 
\l{M1}\ee 
where $b\in\RR^D$ and the matrix $M=A\Lambda$ is the composition of a {\it 
dilatation}
\be 
q\longrightarrow Aq,\qquad A\in\RR, 
\l{zummoloa15}\ee 
and a {\it rotation} 
\be 
q\longrightarrow\Lambda q, 
\l{zummoloa16}\ee 
where $\Lambda\in O(D)$. Similarities are naturally extended to the 
compactified space
$\hat\RR^D=\RR^D\cup\{\infty\}$. A similarity maps $\infty$ to itself. 
 
Let us consider the hyperplane 
\be 
P(a,t)=\{\,q\in\RR^D|q\cdot a=t,a\in\RR^D,t\in\RR\}. 
\l{zummoloa17}\ee 
The {\it reflection} with respect to $P(a,t)$ is given by 
\be 
f(q)=q-2\left(q\cdot a-t\over a\cdot a\right)a. 
\l{reflection}\ee 
Let us set 
\be 
r^2=q_1^2+\cdots+q_D^2. 
\l{zummoloa18}\ee 
The last generator of the M\"obius group is the {\it inversion} or reflection 
in the unit sphere $\SS^{D-1}$. If $q\ne0$ 
\be 
q\longrightarrow q^*={q\over r^2}, 
\l{bosse3}\ee 
otherwise 
\be 
0\longrightarrow\infty,\qquad\infty\longrightarrow0. 
\l{inversion}\ee 
 
The M\"obius group $M(\hat{\RR}^D)$ is defined as the set of transformations 
generated by all similarities together with the inversion. Actually, an 
arbitrary M\"obius transformation is the composition of a number of 
reflections and inversions. Furthermore, a M\"obius transformation is 
conformal with respect to the euclidean metric and a theorem due to Liouville 
states that the conformal group and $M(\hat{\RR}^D)$ actually coincide for 
$D>2$. 
 
\subsection{Translations and dilatations}\label{transldilat} 
 
We now begin to study the properties of $\Z(q^b;q^a)$ when $q^b$ and 
$q^a$ are related by dilatations and translations. 
Let us start by noticing that if $B$ and $C$ are arbitrary constant vectors, 
then from (\ref{cociclo3}) we have 
\be 
(q+B+C;q)=(q+B+C;q+B)+(q+B;q)=(q+B+C;q+C)+(q+C;q), 
\l{eqq2}\ee 
so that 
\be 
(q+B+C;q+B)-(q+B+C;q+C)=(q+C;q)-(q+B;q), 
\l{eqq3}\ee 
where $(q+B)_k=q_k+B_k$, $k=1,\ldots, D$. We will show that the unique 
solution of (\ref{eqq3}) is 
\be 
(q+B;q)=F(q+B)-F(q), 
\l{eqqq3}\ee 
where $F$ is an arbitrary function of $q$. Pick $j\in[1,D]$ and let $B_j$ and 
$C_j$ be the only non--vanishing components of $B$ and $C$ 
\be 
B=(0,\ldots,B_j,0,\ldots),\qquad C=(0,\ldots,C_j,0,\ldots). 
\l{ecchilo}\ee 
Let us set 
\be 
f(B,q)=(q+B;q), 
\l{zummoloa19}\ee 
with $B$ given by (\ref{ecchilo}). Eq.(\ref{eqq3}) reads 
\be 
f(C,q+B)-f(B,q+C)=f(C,q)-f(B,q). 
\l{eqq3bisse}\ee 
Taking the derivative of both sides of Eq.(\ref{eqq3bisse}) with respect to 
$B_j$, we get
\be 
\partial_{q_j}f(C,q+B)-\partial_{B_j}f(B,q+C)=-\partial_{B_j}f(B,q). 
\l{eqq4}\ee 
By (\ref{inparticolareb}) 
\be 
f(B,q)=\sum_{n=1}^\infty c_n(q)B_j^n. 
\l{eqq5}\ee 
Plugging this expression into Eq.(\ref{eqq4}), we find 
\be 
\sum_{n=1}^{\infty}\partial_{q_j}c_n(q+B)C^n_j-\sum_{n=1}^{\infty}n 
c_n(q+C)B_j^{n-1}=-\sum_{n=1}^{\infty}nc_n(q)B_j^{n-1}. 
\l{eqq6}\ee 
Furthermore, expanding $c_n(q+B)$ and $c_n(q+C)$, it follows by 
Eq.(\ref{eqq6}) that
\be 
{1\over(m-1)!}\partial_{q_j}^{m}c_n(q)={m\over n!}\partial_{q_j}^{n}c_m(q), 
\l{eqq7}\ee 
that for $n=1$ reads 
\be 
{1\over m!}\partial_{q_j}^{m}c_1(q)=\partial_{q_j}c_m(q). 
\l{eqq8}\ee 
It follows that 
\be 
\partial_{q_j}f(B,q)=\sum_{n=1}^{\infty}\partial_{q_j}c_n(q)B_j^n= 
\sum_{n=1}^\infty{1\over n!}\partial_{q_j}^nc_1(q)B_j^n=c_1(q+B)-c_1(q), 
\l{eqq9}\ee 
which upon integration on $q_j$ yields 
\be 
f(B,q)=c(q+B)-c(q)+g(B,\hat q), 
\l{eqq10}\ee 
with $\hat q$ denoting all the variables other than $q_j$. Moreover, by 
(\ref{inparticolareb}) $g(0,\hat{q})=0$. Let us show that $g(B,\hat{q})$ is 
identically vanishing. By (\ref{cociclo3}) 
\be 
f(B+C,q)=(q+B+C;q)=(q+B+C;q+B)+(q+B;q)=f(C,q+B)+f(B,q), 
\l{eqq10a}\ee 
that by (\ref{eqq10}) implies 
\be 
g(B+C,\hat{q})=g(B,\hat{q})+g(C,\hat{q}), 
\l{eqq10b}\ee 
that is $g(B,\hat{q})=g'(0,\hat{q})B_j$. However, by (\ref{eqq5})(\ref{eqq9}) 
and (\ref{eqq10}) 
\be 
\partial_{B_j}f(B,q)=\partial_{q_j}c(q+B)+g'(0,\hat{q})= 
c_1(q+B)+g'(0,\hat{q})=\sum_{n=1}^{\infty}nc_n(q)B_j^{n-1}. 
\l{eqq10d}\ee 
Then, setting $B_j=0$, we find from the last equality that $g'(0,\hat{q})=0$. 
Finally, we are left with 
\be 
f(B,q)=c(q+B)-c(q). 
\l{eqq10e}\ee
{}From this equation it is then possible to derive Eq.(\ref{eqqq3}). The 
technical detailes are reported in Appendix A. Related reasonings, reported in
Appendix A, show that
\be 
(Aq;q)=A^2F(Aq)-F(q), 
\l{eqq15b}\ee 
where now 
\be 
F(0)=0,
\l{Finzeroezero}\ee 
with $F$ the same function appearing in Eq.(\ref{eqqq3})

\subsection{Rotations} 
 
Let us consider $(\Lambda q;q),$ where $\Lambda\in O(D)$. First of all, if
$q^b=\Lambda q^a$, we see that $(p^b|p^a)=1$, because $\Lambda^t\Lambda=
\Lambda\Lambda^t={\bf 1}_D$. Hence, by (\ref{cociclo3}) 
$$ 
(\Lambda(q+B);q+B)=(\Lambda q+\Lambda B;q+B)= (\Lambda q+\Lambda B;\Lambda 
q)+(\Lambda q;q+B)= 
$$ 
\be 
(\Lambda q+\Lambda B;\Lambda q)+(\Lambda q;q)+(q;q+B), 
\l{rot1}\ee 
which implies 
\be 
(\Lambda(q+B);q+B)-(\Lambda q;q)=F(\Lambda(q+B))-F(\Lambda q)+F(q)-F(q+B). 
\l{rot1a}\ee 
Therefore, $(\Lambda q;q)=F(\Lambda q)-F(q)+c$. However, since $(\Lambda q;q)$ 
evaluated at $q=0$ cannot depend on $\Lambda$, we have $(\Lambda q;q)_{q=0}= 
(q;q)_{q=0}=(q;q)=0$. Then 
\be 
(\Lambda q;q)=F(\Lambda q)-F(q). 
\l{rot1b}\ee 
 
\subsection{Inversion} 
 
Let us consider the inversion $q^*(q)$ (\ref{bosse3}). The Jacobian matrix of 
this mapping is given by 
\be 
J_{kl}={\partial q^*_l\over\partial q_k}={\partial\over\partial q_k} 
\left({q_l\over r^2}\right)={\delta_{kl}\over r^2}-2{q_kq_l\over r^4}. 
\l{lappona1}\ee 
Then 
\be 
(J^tJ)_{jk}=(J^2)_{jk}=\sum_{l=1}^DJ_{jl}J_{lk}=\sum_{l=1}^D\left({\delta_{jl} 
\over r^2}-2{q_j q_l\over r^4}\right)\left({\delta_{lk}\over r^2}-2{q_lq_k\over
r^4}\right)={\delta_{jk}\over r^4}, 
\l{lappona2}\ee 
which implies 
\be 
(p|p_*)=(p_*|p)^{-1}={\sum_kp_k^2\over\sum_k {p_*}_k^2}={p_*^tJ^tJp_*\over 
p_*^tp_*}={1\over r^4}. 
\l{jacobinver}\ee 
Note that $q^*$ is involutive since 
\be 
r_*^2=\sum_{k=1}^Dq^*_kq^*_k={1\over r^4}\sum_{k=1}^Dq_kq_k={1\over r^2}, 
\l{rdieffe}\ee 
and therefore 
\be 
(q^*)^*_k={q^*_k\over r_*^2}={q_k\over r^2r_*^2}=q_k. 
\l{involutiva}\ee 
Observe that since rotations leave $r$ invariant, we have 
\be 
(\Lambda q)^*_j={(\Lambda q)_j\over r_{\Lambda}^2}= {\Lambda_{jk}q_k\over 
r^2}=\Lambda_{jk}q^*_k=(\Lambda q^*)_j. 
\l{commutarotazione}\ee 
Finally, we recognize the following behaviour under dilatations 
\be 
(Aq)^*_j={(Aq)_j\over r_A^2}={Aq_j\over A^2r^2}=A^{-1}{q_j\over r^2}=A^{-1} 
q^*_j,
\l{dilatazioni}\ee 
where $r_A^2=\sum_{k=1}^D Aq_k Aq_k=A^2 r^2$. By (\ref{jacobinver}) and 
(\ref{involutiva}) 
\be 
(q^*;q)=-(p|p^*)(q;q^*)=-{1\over r^4}((q^*)^*;q^*), 
\l{iappala}\ee 
which implies that $(q^*;q)$ vanishes when evaluated at any $q_0$ solution of 
$q^*=q$, that is 
\be 
(q^*;q)|_{q=q_0}=0. 
\l{eppercheno}\ee
{}From this one derives the following result
\be 
(q^*;q)={1\over r^4}F(q^*)-F(q),
\l{inver1e}\ee 
whose proof is reported in Appendix A.

\subsection{Fixing the coboundary}\l{coboundary} 
 
Let us denote by $\gamma(q)$ a M\"obius transformation of $q$. By 
(\ref{eqqq3})(\ref{eqq15b})(\ref{rot1b}) and (\ref{inver1e}) we have 
\be 
(\gamma(q);q)=(p|p^\gamma)F(\gamma(q))-F(q). 
\l{egiaegia}\ee 
Given a function $f(q)$, we have that if $(f(q);q)$ satisfies the cocycle 
condition (\ref{cociclo3}), then this is still satisfied under the 
substitution 
\be 
(f(q);q)\longrightarrow (f(q);q)+(p|p^f)G(f(q))-G(q), 
\l{ftpfenderstratocaster}\ee 
where $G$ has to satisfy the condition $G(0)=0$. This condition is a 
consequence of the fact that $(Aq;q)$ evaluated at $q=0$ is independent of 
$A$, so that it vanishes at 
$q=0$. Therefore, if $(Aq;q)$ satisfies (\ref{cociclo3}), then also 
$(Aq;q)+A^2G(Aq)-G(q)$ should vanish at $q=0$, implying that $G(0)(A^2-1)=0$, 
that is $G(0)=0$. The term $(p|p^f)G(f(q))-G(q)$ can be seen as a coboundary 
term. We now show how the coboundary ambiguity (\ref{ftpfenderstratocaster}) 
is fixed. First of all observe that (\ref{is1}) implies 
\be 
\S_0^0(q^0)=\S_0(q), 
\l{is1perilvuoto}\ee 
where $\S_0^0$ denotes the reduced action associated to the $\W^0\equiv0$ 
state. On the other hand, by (\ref{ddazzoyyyxxaa10bbbb}) we have that the 
equation of motion for 
$\S_0(q)$ we are looking for is 
\be 
\Z(q^0;q)=\W(q). 
\l{ordinariooperstradamasiamoinitalia}\ee 
Comparing (\ref{ordinariooperstradamasiamoinitalia}) with (\ref{traslazione}) 
and (\ref{is1perilvuoto}) we see that a necessary condition to satisfy 
(\ref{traslazione}) is that $(q^0;q)$ depends only on the first and higher 
derivatives of $q^0$. In fact, by (\ref{is1perilvuoto}) we have 
$q^0={\S_0^0}^{ -1}\circ\S_0(q)$, that is $q^0$ is a functional of $\S_0$. 
Therefore, a possible dependence of $\Z(q^0;q)$ on $q^0$ itself, would imply 
that (\ref{ordinariooperstradamasiamoinitalia}) has the form 
$\F(\S_0,\nabla\S_0, 
\Delta\S_0,\ldots)=0$ rather than (\ref{traslazione}). Therefore, the only 
possibility is that the function $F$ in (\ref{egiaegia}) be vanishing 
\be 
F=0. 
\l{inver1n}\ee 
 
Therefore, we arrived at the following basic result 
 
\vspace{.333cm} 
 
\noindent 
{\it Eq.(\ref{traslazione}) and the cocycle condition (\ref{cociclo3}) imply 
that $(q^a;q^b)$ vanishes when $q^a$ and $q^b$ are related by a M\"obius 
transformation, that is} 
\be 
(q+B;q)=0, 
\l{eqq18}\ee 
\be 
(Aq;q)=0, 
\l{eqq17}\ee 
\be 
(\Lambda q;q)=0, 
\l{rotazioni}\ee 
\be 
(q^*;q)=0. 
\l{inversione2}\ee 
 
\vspace{.333cm} 

The above equations are equivalent to $(\gamma(q);q)=0$. Furthermore, by 
(\ref{cociclo3}) we have 
\be 
(\gamma(q^a);q^b)=(q^a;q^b),\qquad (q^a;\gamma(q^b))=(p^{\gamma(b)}|p^b) 
(q^a;q^b).
\l{cs}\ee 
Let us consider the Jacobian factor $(p^{\gamma(b)}|p^b)$. First of all 
observe that the M\"obius transformation is conformal with respect to the 
Euclidean metric. Namely, we have 
\be 
ds^2=\sum_{j=1}^D d\gamma(q)_jd\gamma(q)_j=\sum_{j,k,l=1}^D{\partial\gamma(q)_j
\over\partial q_k}{\partial\gamma(q)_j\over\partial q_l}dq_kdq_l=\sum_{j=1}^D 
e^{\phi_\gamma(q)}dq_jdq_j. 
\l{conf1}\ee 
Therefore 
\be 
(p^{\gamma(b)}|p^b)=e^{-\phi_{\gamma}(q^b)}. 
\l{conf2}\ee 
Note that in the case of translations and rotations the conformal re--scaling 
is the identity. For dilatations $\exp\phi^A=A^2$, whereas for the inversion 
$\exp\phi^*=r^{-4}$. 
 
Note that the above conformal structure arises by setting 
$\S_0^v(q^v)=\S_0(q)$. Let us make clear that this is not an assumption. Any 
transformation we choose other than 
$\S_0^v(q^v)=\S_0(q)$ would yield the same results. In particular, the
absence of assumptions in setting $\S_0^v(q^v)=\S_0(q)$ results from the fact 
that $q$ and $q^v$ represent the spatial coordinates in their own systems. So, 
$\S_0^v(q^v)=\S_0(q)$ can be seen just as the simplest way to set the 
coordinate transformations from the system with reduced action $\S^v_0$ (since 
physics is determined by the functional structure of 
$\S^v_0$, we can denote the coordinate as we like) to the one with reduced
action
$\S_0$. Nevertheless, there is a hidden apparently ``innocuous" assumption:
that the position $\S_0^v (q^v)=\S_0(q)$ actually makes sense. This is not the 
case in CM, as for the free particle of vanishing energy we have 
$\S_0(q)=cnst$. In this case the above position does not make sense. Requiring 
that this is well--defined for any system is essentially the same as imposing 
the EP. However, on the one hand we have seen that the existence of the 
transformation implies the conformal structure, on the other we will see that 
the EP, and therefore existence of the transformation, implies QM. Thus, we 
have that the M\"obius group, that for $D 
\geq3$ coincides with the conformal group, is intimately related to QM itself. 
 
\mysection{The Schr\"odinger equation}\l{lascr} 
 
In this section, we will derive the quantum Hamilton--Jacobi equation in 
$D$ dimensions and then show that the latter is equivalent to the stationary 
Schr\"odinger equation. 
 
Let us start with the Quantum Stationary HJ Equation (QSHJE) in one dimension 
\be 
{1\over2m}\left({\partial\S_0\over\partial q}\right)^2+V(q)-E+{\hbar^2\over4m} 
\{\S_0,q\}=0. 
\l{aa10bbbxxxb}\ee 
This equation was univocally derived from the EP in \cite{1}\cite{2}. After
publishing
\cite{1}, the authors became aware that this equation was assumed in
\cite{Floyd} as a starting point to formulate a trajectory interpretation of
QM. In particular, Floyd \cite{Floyd} introduced the concept of trajectories 
by using Jacobi's theorem according to which 
\be 
t-t_0={\partial\S_0\over\partial E}, 
\l{Floydtime}\ee 
{}from which one sees that the conjugate momentum $p=\partial_q\S_0$ does not 
in general correspond to the mechanical one, that is $p\ne m\dot q$. This is a 
basic difference with respect to Bohm's theory 
\cite{Bohm}--\cite{recentproposal}. Furthermore, Floyd 
noted that Bohm's assumption $\psi=Re^{{i\over\hbar}\hat\S_0}$ does not work 
in this case \cite{Floyd}. Apparently one may infer that (\ref{aa10bbbxxxb}) 
is equivalent to the standard version of the quantum stationary HJ equation 
\be 
{1\over2m}\left({\partial\hat\S_0\over\partial q}\right)^2+V(q)-E-{\hbar^2\over
2m}{\partial_q^2R\over R}=0, 
\l{aa10bbbxxxbdibohm}\ee 
\be 
\partial_q (R^2\partial_q\hat\S_0)=0. 
\l{lacontinua}\ee 
In fact, solving (\ref{lacontinua}) would give 
\be 
R={c\over\sqrt{\partial_q\hat S_0}}, 
\l{reds}\ee 
which is equivalent to 
\be 
\{\hat\S_0,q\}=-2{\partial_q^2R\over R}, 
\l{abced}\ee 
so that $\hat\S_0$ would satisfy the same equation as $\S_0$. Nevertheless, 
there is a problem in the above derivation. Namely, in Bohm's assumption, like 
in the usual formulation of quantum HJ theory, the identification is not 
between a general solution of the Schr\"odinger equation and 
$Re^{{i\over\hbar}\hat\S_0}$, but between the wave--function and 
$Re^{{i\over\hbar}\hat\S_0}$. Thus, suppose that the wave--function describes 
a bound state so that it must be proportional to a real 
function.\footnote{This is a consequence of reality of $\W$ as this implies 
that if 
$\psi$ solves the Schr\"odinger equation, then this is the case also of
$\bar\psi$. If $\bar\psi\not\propto\psi$, then $\psi\bar\psi'-
\psi'\bar\psi=cnst\ne0$, so that $\psi$ is never vanishing. In particular,
if $\psi\in L^2(\RR)$, then $\bar\psi\propto\psi$ (see also sections 14 and 17 
of Ref.\cite{6}).} According to Bohm and the usual approach, this would 
imply\footnote{To be precise, bound states would correspond to 
$\hat\S_0=cnst$ outside the nodes of the wave--function.} 
$$ 
\hat\S_0\; is\; a\; constant\; for\; bound\; states. 
$$ 
This in turn implies rather peculiar properties. For example, quantum 
mechanically the conjugate momentum is vanishing for bound states. This seems
to be an unsatisfactory feature of 
(\ref{aa10bbbxxxbdibohm})(\ref{lacontinua}). To be more precise, 
Eqs.(\ref{aa10bbbxxxbdibohm})(\ref{lacontinua}) are good equations unless one 
forces the identification of $Re^{{i\over\hbar}\hat\S_0}$ with the 
wave--function. In general 
$Re^{{i\over\hbar}\hat\S_0}$ should be identified with a linear combination
of two linearly independent solutions of the Schr\"odinger equation. So, the 
general expression for the wave--function is
\be 
\psi=R\left(A e^{-{i\over\hbar}\S_0}+Be^{{i\over\hbar}\S_0}\right), 
\l{aebbb}\ee
so that since $\bar\psi\propto\psi$ gives 
\be 
|A|=|B|. 
\l{amodulob}\ee 
Thus for $\S_0$ there is no trace of the condition $\hat\S_0=cnst$ one has for
bound states setting $\psi=Re^{{i\over\hbar}\hat\S_0}$. Let us note that 
whereas $\hat\S_0=cnst$ would be consistent in the case of classically 
forbidden regions, as there $\S_0^{cl}=cnst,$\footnote{Note also that having 
$\hat\S_0=cnst$ in the classically forbidden regions would imply a trivial 
trajectory, since $p=0$ there.} problems arise in regions which are not 
classically forbidden, $i.e.$ where $\S_0^{cl}\ne cnst$. For example, in the 
case of the harmonic oscillator, one has $\hat\S_0=cnst$ $\forall q\in\RR$, 
which follows by identifying $R\exp(i\hat\S_0/\hbar)$ with the wave--function, 
while in some region one has $\S_0^{cl}\ne cnst$. As a consequence, while 
quantum mechanically the particle would be at rest, after taking the 
$\hbar\to0$ limit, the particle should start moving. We refer to Holland's 
book \cite{Holland} for an interesting analysis concerning the classical limit 
of the harmonic oscillator in Bohmian theory. 
 
The above analysis can be summarized by the following basic fact 
 
\vspace{.333cm} 
 
\noindent 
{\it If $\hat\S_0$ is the quantum analogue of the reduced action, and 
therefore reduces to the classical one in the $\hbar\to0$ limit, then the 
wave--function cannot be generally identified with $R\exp(i\hat\S_0/\hbar)$. 
In particular, this cannot be the case for bound states, such as the harmonic 
oscillator, in which the wave--function is proportional to a real function 
also in regions which are not classically forbidden.} 
 
\vspace{.333cm} 
 
However, we have seen that if $R\exp(i\hat\S_0/\hbar)$ is not identified with
real solutions of the Schr\"odinger equation, then we have equivalence between 
(\ref{aa10bbbxxxb}) and (\ref{aa10bbbxxxbdibohm})(\ref{lacontinua}). We also 
note that with the formulation (\ref{aa10bbbxxxb}) one directly sees that the 
situation 
$\S_0=cnst$ can never occur. In fact, one has rather stringent conditions 
connected with the existence of $\{\S_0,q\}$, which in turn reflects the basic 
nature of the cocycle condition and therefore of the EP. In this respect we 
recall that existence of $\{\S_0,q\}$ implies that the ratio of two real 
linearly independent solutions of the Schr\"odinger equation must be a local 
self--homeomorphism of the extended real line $\hat\RR=\RR\cup\{\infty\}$. 
This is a basic fact as it implies energy quantization without any assumption 
\cite{5}\cite{6}. 
 
\subsection{Hidden variables, Planck length and holographic principle} 
 
The above remarks are related to the proposal of changing the Bohmian
definition of mechanical momentum $m\dot q$ \cite{recentproposal}. This 
proposal is related to the fact that
(\ref{aa10bbbxxxbdibohm})(\ref{lacontinua}) allow a rearrangement of 
$\hat\S_0$ and $R$. On the other hand, these symmetries are particularly 
evident working directly with Eq.(\ref{aa10bbbxxxb}), which in turn is 
equivalent to the Schwarzian equation 
\be 
\left\{e^{{2i\over\hbar}\S_0},q\right\}=-{4m\over\hbar^2}\W. 
\l{scharziana}\ee 
In fact these symmetries correspond to the invariance of (\ref{scharziana})
under M\"obius transformations of $e^{{2i\over\hbar}\S_0}$. 
 
Eq.(\ref{aa10bbbxxxb}) implies that $\S_0$ can be expressed in the ``canonical
form" \cite{1}\cite{2}\cite{6} 
\be 
e^{{{2i}\over\hbar}\S_0\{\delta\}}=e^{i\alpha}{w+i\bar\ell\over w-i\ell}, 
\l{dfgtp}\ee 
which is equivalent to the one considered by Floyd \cite{Floyd}. Here $\delta= 
\{\alpha,\ell\}$, where $\alpha\in\RR$ and $\ell_1={\rm Re}\,\ell\ne0$, $\ell_2
={\rm Im}\,\ell$ are integration constants, and $w=\psi^D/\psi\in\RR$, where $ 
\psi^D$ and $\psi$ are real linearly independent solutions of the stationary 
Schr\"odinger equation 
\be 
\left[-{\hbar^2\over2m}{\partial^2\over\partial q^2}+V(q)\right]\psi=E\psi. 
\l{yz1xxxx4}\ee 
Observe that the condition $\ell_1\ne0$ is equivalent to having $\S_0\ne cnst$ 
which is a necessary condition to define $\{\S_0,q\}$ in the QSHJE. 
 
A basic feature of (\ref{dfgtp}) is that it explicitly shows the existence of 
M\"obius states \cite{1}--\cite{6}, called microstates by Floyd \cite{Floyd}. 
In particular, the constants $\ell_1$ and $\ell_2$ correspond, together with 
$\alpha$, to the initial conditions of Eq.(\ref{aa10bbbxxxb}). These initial 
conditions do not appear in the Schr\"odinger equation, so that $\ell_1$ and 
$\ell_2$ can be seen as a sort of hidden variables. Their role is quite basic. 
In particular, it has been shown in \cite{4}\cite{6} that in order to have a 
well--defined classical limit, $\ell$ should depend on fundamental lengths 
which in turn should depend on $\hbar$. This dependence arises in considering 
the 
$E\to0$ and $\hbar\to0$ limits. In particular, let us consider the conjugate 
momentum in the case of the free particle with energy $E$ 
\be 
p_E=\pm{\hbar(\ell_E+\bar\ell_E)\over2|k^{-1}\sin kq-i\ell_E\cos kq|^2}, 
\l{dajjj}\ee 
where $k=\sqrt{2mE}/\hbar$. The first condition is that in the $\hbar 
\to0$ limit the conjugate momentum reduces to the classical one 
\be 
\lim_{\hbar\longrightarrow0}p_E=\pm\sqrt{2mE}. 
\l{bos1S11b}\ee 
On the other hand, we should also have 
\be 
\lim_{E\longrightarrow0}p_E=p_0=\pm{\hbar(\ell_0+\bar\ell_0)\over2 
|q-i\ell_0|^2}. 
\l{bisCS11b}\ee 
 
Eqs.(\ref{dajjj})(\ref{bos1S11b}) show that, due to the factor $\hbar$ in 
$\cos kq$, the quantity $\ell_E$ should depend on $E$. Let us set 
\be 
\ell_E=k^{-1}f(E,\hbar)+\lambda_E, 
\l{prova4}\ee 
where, since $\lambda_E$ is still arbitrary, we can choose the dimensionless 
function $f$ to be real. By (\ref{dajjj}) we have 
\be 
p_E=\pm{\sqrt{2mE}f(E,\hbar)+mE(\lambda_E+\bar\lambda_E)/\hbar\over 
\left|e^{ikq}+(f(E,\hbar)-1+\lambda_Ek)\cos kq\right|^2}. 
\l{prova4cc}\ee 
Note that if one ignores $\lambda_E$ and sets $\lambda_E=0$, then by 
(\ref{bos1S11b})
\be 
\lim_{\hbar\longrightarrow0}f(E,\hbar)=1. 
\l{prova5}\ee 
We now consider the properties that $\lambda_E$ and $f$ should have in order 
that (\ref{prova5}) be satisfied in the physical case in which $\lambda_E$ is 
arbitrary but for the condition ${\rm Re}\,\ell_E\ne0$, as required by the 
existence of the QSHJE. First of all note that cancellation of the divergent 
term $E^{-1/2}$ in 
\be 
p_E\;{}_{\stackrel{\sim}{E\longrightarrow0}}\pm{2\hbar^2(2mE)^{-1/2}f(E, 
\hbar)+\hbar(\lambda_E+\bar\lambda_E)\over2|q-i\hbar(2mE)^{-1/2}f(E,\hbar)-i 
\lambda_E|^2}, 
\l{prova3truciolo}\ee 
yields 
\be 
\lim_{E\longrightarrow0}E^{-1/2}f(E,\hbar)=0, 
\l{prova6}\ee 
so that $k$ must enter in the expression of $f(E,\hbar)$. Since $f$ is a 
dimensionless constant, we need at least one more constant with the dimension 
of length. Two fundamental lengths one can consider are the Compton length 
$\lambda_c={\hbar/mc}$, and the Planck length $\lambda_p=\sqrt{\hbar G/c^3}$. 
Two dimensionless quantities depending on $E$ are 
\be 
x_c=k\lambda_c=\sqrt{2E\over mc^2}, 
\l{kComptlength}\ee 
and 
\be 
x_p=k\lambda_p=\sqrt{2mEG\over\hbar c^3}. 
\l{kPlancklength}\ee 
On the other hand, since $x_c$ does not depend on $\hbar$ it cannot be used to 
satisfy (\ref{prova5}), so that it is natural to consider $f$ as a function of 
$x_p$. Let us set 
\be 
f(E,\hbar)=e^{-\alpha(x_p^{-1})}, 
\l{prova9}\ee 
where $\alpha(x_p^{-1})=\sum_{k\geq 1}\alpha_kx_p^{-k}$. The conditions 
(\ref{prova5})(\ref{prova6}) correspond to conditions on the coefficients 
$\alpha_k$. In order to consider the structure of $\lambda_E$, we note that 
although $e^{-\alpha(x_p^{-1})}$ cancelled the $E^{-1/2}$ divergent term, we 
still have some conditions to be satisfied. To see this note that 
\be 
p_E=\pm{\sqrt{2mE}e^{-\alpha(x_p^{-1})}+mE(\lambda_E+\bar\lambda_E)/\hbar 
\over\left|e^{ikq}+(e^{-\alpha(x_p^{-1})}-1+k\lambda_E)\cos kq\right|^2}, 
\l{prova20}\ee 
so that (\ref{bos1S11b}) implies 
\be 
\lim_{\hbar\longrightarrow0}{\lambda_E\over\hbar}=0. 
\l{prova21}\ee 
To discuss this limit, we first note that 
\be 
p_E=\pm{2\hbar k^{-1}e^{-\alpha(x_p^{-1})}+\hbar(\lambda_E+\bar\lambda_E)\over 
2\left|k^{-1}\sin kq-i\left(k^{-1}e^{-\alpha(x_p^{-1})}+\lambda_E\right)\cos 
kq\right|^2}. 
\l{prova21bbv}\ee 
So that, since $\lim_{E\longrightarrow0}k^{-1}e^{-\alpha(x_p^{-1})}=0$, by 
(\ref{bisCS11b}) and (\ref{prova21bbv}) we have 
\be 
\lambda_0= 
\lim_{E\longrightarrow0}{\lambda_E}=\lim_{E\longrightarrow0}{\ell_E}=\ell_0. 
\l{prova22}\ee 
Let us now consider the limit 
\be 
\lim_{\hbar\longrightarrow0}p_0=0. 
\l{prova23}\ee 
First of all note that, since 
\be 
p_0=\pm{\hbar(\ell_0+\bar\ell_0)\over2|q^0-i\ell_0|^2}, 
\l{prova24}\ee 
we have that the effect on $p_0$ of a shift of ${\rm Im}\,\ell_0$ is 
equivalent to a shift of the coordinate. Therefore, in considering 
(\ref{prova23}) we can set ${\rm Im}\,\ell_0=0$ and distinguish the cases 
$q^0\ne0$ and $q^0=0$. Note that as we always have ${\rm Re}\,\ell_0\ne0$, it 
follows that the denominator in the right hand side of (\ref{prova24}) is 
never vanishing. Let us define 
$\gamma$ by 
\be 
{\rm Re}\,\ell_0{}_{\stackrel{\sim}{\hbar\longrightarrow0}}\hbar^{\gamma}. 
\l{ioq990p}\ee 
We have 
\be 
p_0{}_{\stackrel{\sim}{\hbar\longrightarrow0}}\left\{\begin{array}{ll}\hbar^{ 
\gamma+1}, & q_0\ne0,\\ \hbar^{1-\gamma},& q_0=0,\end{array}\right. 
\l{Vu1fy9}\ee 
and by (\ref{prova23}) 
\be 
-1<\gamma<1. 
\l{prova25}\ee 
A constant length having powers of $\hbar$ can be constructed by means of 
$\lambda_c$ and $\lambda_p$. We also note that a constant length which is 
independent of $\hbar$ is provided by $\lambda_e=e^2/mc^2$ where $e$ is the 
electric charge. Thus $\ell_0$ can be considered as a suitable function of 
$\lambda_c$, $\lambda_p$ and $\lambda_e$ satisfying the constraint 
(\ref{prova25}). 
 
The above investigation indicates that a natural way to express $\lambda_E$ is 
given by
\be 
\lambda_E=e^{-\beta(x_p)}\lambda_0, 
\l{prova26}\ee 
where $\beta(x_p)=\sum_{k\geq 1}\beta_k x_p^k$. Any possible choice of $\beta 
(x_p)$ should satisfy the conditions (\ref{prova21}) and (\ref{prova22}). For 
example, for the modulus $\ell_E$ built with $\beta(x_p)=\beta_1 x_p$, one 
should have $\beta_1>0$. 
 
Summarizing, by (\ref{prova4})(\ref{prova9})(\ref{prova22}) and 
(\ref{prova26}) we have
\be 
\ell_E=k^{-1}e^{-\alpha(x_p^{-1})}+e^{-\beta(x_p)}\ell_0, 
\l{prova27}\ee 
where 
$\ell_0=\ell_0(\lambda_c,\lambda_p,\lambda_e)$, and for the conjugate 
momentum of the state $\W=-E$ we have 
\be 
p_E=\pm{2k^{-1}\hbar e^{-\alpha(x_p^{-1})}+\hbar e^{-\beta(x_p)}(\ell_0+\bar 
\ell_0)\over2\left|k^{-1}\sin kq-i\left(k^{-1}e^{-\alpha(x_p^{-1})}+e^{-\beta 
(x_p)}\ell_0\right)\cos kq\right|^2}. 
\l{prova21bbvxx}\ee 
 
We stress that the appearance of the Planck length is strictly related to 
$p$--$q$ duality and to the existence of the Legendre transformation of
$\S_0$ for any state. This $p$--$q$ duality has a counterpart in the
$\psi^D$--$\psi$ duality \cite{1}--\cite{6} which sets a length scale that
already appears in considering linear combinations of $\psi^{D^0}=q^0$ and 
$\psi^0=1$. This aspect is related to the fact that we always have $\S_0\ne 
cnst$ and $\S_0\not\propto q+cnst$, so that also for the states 
$\W^0$ and $\W=-E$ one has a non--constant conjugate momentum.
In particular, the Planck length naturally emerges in considering 
$\lim_{E\to0}p_E=p_0$, together with the analysis of the 
$\hbar\longrightarrow0$ limit of both $p_E$ and $p_0$. 
As a result the Compton length and $\lambda_e$ appear as well. 
 
We also note that in \cite{2}\cite{6} it has been shown that the 
wave--function remains invariant under suitable transformations of $\alpha$ 
and 
$\ell$. These transformations constitute the basic symmetry group of the 
wave--function. To see this we consider the case of the wave--function 
$\psi_E$ corresponding to a state of energy $E$. Since $\psi_E$ solves the 
Schr\"odinger equation, for any fixed set of integration constants $\alpha$
and $\ell$, there are coefficients $A$ and $B$ such that 
\be 
\psi_E\{\delta\}={1\over\sqrt{\S_0'\{\delta\}}}\left(A e^{-{i\over\hbar} 
\S_0\{\delta\}}+Be^{{i\over\hbar}\S_0\{\delta\}}\right). 
\l{popcaSCHR}\ee 
Performing a transformation of the moduli 
$\delta\to\delta'=\{\alpha',\ell'\}$, we have (we refer to \cite{2}\cite{6} 
for notation) 
\be 
\psi_E\{\delta'\}=\left({2i\over\tilde\mu\hbar\partial_q\gamma_{\S_0}}\right)^{ 
1/2}\left[A\tilde d+B\tilde b+(A\tilde c+B\tilde a)\gamma_{\S_0}\right]. 
\l{ppssid}\ee 
Requiring that $\psi_E\{\delta\}$ remains unchanged up to some multiplicative 
constant $c$, that is 
\be 
\psi_E\{\delta\}\longrightarrow\psi_E\{\delta'\}=c\psi_E\{\delta\}, 
\l{ccooss}\ee 
we have by (\ref{ppssid}) 
\be 
A^2\bar{\tilde b}+AB\tilde a=AB\bar{\tilde a}+B^2\tilde b. 
\l{Kj9}\ee 
This defines the symmetry group of the wave--function. Thus, we have seen that 
there are hidden variables depending on the Planck length and that these can 
be suitably changed without any effect on the wave--function. Therefore, we 
can say that there is a sort of information loss in considering the 
wave--function. So, the probabilistic interpretation of the wave--function 
seems due to our ignorance about Planck scale physics. 
 
\vspace{.333cm} 
 
The above analysis can be summarized as follows 
 
\begin{itemize} 
\item[{\bf 1.}]{QM follows from the EP\cite{1}--\cite{6}. The formulation is 
strictly related to $p$--$q$ duality, which in turn is a consequence of the 
involutive nature of the Legendre transformation. In this context QM is 
described in terms of trajectories where, according to Floyd \cite{Floyd}, 
time parameterization is defined by Jacobi's theorem.} 
\item[{\bf 2.}]{The theory shows the existence of M\"obius states 
\cite{1}--\cite{6}, called microstates by Floyd \cite{Floyd}. These states 
cannot be seen in the framework of ordinary QM. In particular, these states 
appear in the context of the quantum HJ equation whose initial conditions 
depend on the Planck length \cite{4}\cite{6}. Furthermore, from the symmetries 
of the wave--function under change of hidden variables, we explicitly see that 
there are equivalence classes of the moduli $\delta$ which correspond to the 
same wave--function \cite{2}\cite{6}.} 
\item[{\bf 3.}]{The role of $p$--$q$ duality is a fundamental one. In fact, this
reflects in the appearance in the formulation of a pair of real linearly 
independent solutions of the Schr\"odinger equation. So there is a 
$\psi^D$--$\psi$ duality \cite{1}--\cite{6} which reflects the basic M\"obius 
symmetry and therefore the existence of M\"obius states. This directly shows 
that in considering solutions of the basic Schr\"odinger equation 
$\partial_q^2\psi^0=0$, one has to introduce a length to consider linear 
combinations of $\psi^{D^0}=q^0$ and $\psi^0=1$. Since in this case $\W\equiv 
V- E=0$, the Schr\"odinger problem does not provide any scale, so that we are 
forced to introduce a universal length.} 
\item[{\bf 4.}]{Implementation of the EP implies that the trivializing map, 
expressed as the M\"obius transform of $\psi^D/\psi$, must be a local 
self--homeomorphism of $\hat\RR$ \cite{1}--\cite{6}. This in turn implies that 
for suitable $\W$'s the corresponding Schr\"odinger equation must admit an 
$L^2(\RR)$ solution \cite{5}\cite{6}. This implies that the EP itself implies
energy quantization. So basic facts of QM, such as tunnelling and energy 
quantization, are derived without axiomatic assumptions concerning the 
interpretation of the wave--function. Furthermore, the appearance of the
$L^2(\RR)$ condition shows that the Hilbert space structure starts emerging.} 
\end{itemize} 
 
The above shortly summarizes some of the main aspects of the theory. In this 
context we note that the appearance of Planck length in hidden variables has 
been recently advocated by 't Hooft \cite{tHooft}. 't Hooft argues that such 
hidden variables must play a role in the implementation of the holographic 
principle \cite{holo}. In 't Hooft's paper it is also argued that due to 
information loss, Planck scale degrees of freedom must be combined into 
equivalence classes. The presence of equivalence classes moduli $\delta$, 
corresponding to symmetries of the wave--function, seems to be a possible 
framework for 't Hooft's proposal\footnote{We observe that in an interesting 
paper, Floyd has recently considered related issues \cite{FloydtHooft}.}. 
 
\subsection{The higher dimensional case}\l{uccellagione} 
 
Let us now consider the problem of finding the equation for $\S_0$ in the 
higher dimensional case. To this end, let us first consider a potential of the
form 
\be 
V(q)=\sum_{k=1}^DV_k(q_k), 
\l{ftpocc}\ee 
so that 
\be 
\W(q)=\sum_{k=1}^D\W_k(q_k), 
\l{ftpo}\ee 
where 
\be 
\W_k(q_k)=V_k(q_k)-E_k. 
\l{oi3x}\ee 
In this case, since 
\be 
{1\over2m}({\partial_k\S_{0,k}(q_k)})^2+\W_k(q_k)+Q_k(q_k)=0,\quad 
k=1,\ldots,D,
\l{saa10bbbb}\ee 
we have 
\be 
{1\over2m}\sum_{k=1}^D({\partial_k\S_0(q)})^2+\W(q)+Q(q)=0, 
\l{aa10bbbbb}\ee 
where 
\be 
\S_0(q)=\sum_{k=1}^D\S_{0,k}(q_k),\qquad Q(q)=\sum_{k=1}^DQ_k(q_k), 
\l{oiwn}\ee 
and 
\be 
Q_k(q_k)={\hbar^2\over 4m}\{\S_{0,k}(q_k),q_k\}. 
\l{oiwn2}\ee 
Note that by (\ref{azzoyyyxxaa10bbbb})(\ref{azzo2yyyxxaa10bbbb})(\ref{eqq18}) 
and (\ref{rotazioni}), both $\W(q)$ and $Q(q)$ are invariant under rotations 
and  translations\footnote{It is worth stressing that these transformations 
are not a  symmetry of the physical system as in general the functional 
structures change, that is  
$\tilde \W(x)\ne \W(x)$, $\tilde Q(x)\ne Q(x)$. Thus, 
Eq.(\ref{rota}) should not be confused with true symmetries, {\it e.g.} 
invariance of the potential under  
rotations, expressed as $\W(\Lambda q)=\W(q)$.} 
\be 
\tilde\W(\tilde q)=\W(q(\tilde q)),\qquad\qquad 
\tilde Q(\tilde q)=Q(q(\tilde q)),\qquad\qquad\tilde{q}=\Lambda q+b. 
\l{rota}\ee 
However, observe that 
\be 
\tilde Q(\tilde q)=Q(q(\tilde q))={\hbar^2\over4m}\sum_{k=1}^D\{\S_0(q),
q_k\}={\hbar^2\over4m}\sum_{k=1}^D\{\tilde\S_0(\tilde q),q_k(\tilde q)\}\ne 
{\hbar^2\over4m}\sum_{k=1}^D\{\tilde\S_0(\tilde q),\tilde q_k\}. 
\l{oiwn3}\ee 
This means that expressing $Q(q)$ in terms of sums of Schwarzian derivatives 
does not provide a convenient, {\it i.e.} covariant, formulation. In the 
following, we will express the quantum potential $Q$, and consequently the 
QSHJE in a way that makes this invariance manifest. First of all, note that 
any 
$Q_k$ can be written as 
\be 
Q_k(q_k)=-{\hbar^2\over2m}{\Delta_k R_k\over R_k},\qquad\qquad 
{\partial_k(R_k^2\partial_k\S_{0,k}(q_k))}=0. 
\l{bella1}\ee 
In fact, as we have seen above, since the implementation of the EP implies
that $\S_0$ is never a constant, we have (\ref{abced}). Therefore, we have 
\be 
Q(q)=\sum_{k=1}^DQ_k(q_k)=-{\hbar^2\over2m}\sum_{k=1}^D{\Delta_kR_k 
\over R_k}=-{\hbar^2\over2m}{\Delta R\over R}, 
\l{oleole}\ee 
where $R(q)=\prod_{k=1}^DR_k(q_k)$ satisfies the continuity equation 
\be 
\sum_{k=1}^D{\partial_k(R^2\partial_k\S_0)}=0, 
\l{bella2}\ee 
where $\S_0(q)=\sum_{k=1}^D\S_{0,k}(q_k)$. Now consider the following basic 
identity, which generalizes the one--dimensional version \cite{1}--\cite{6} 
\be 
\alpha^2(\nabla\S_0)^2={\Delta(Re^{\alpha\S_0})\over Re^{\alpha 
\S_0}}-{\Delta R\over R}-{\alpha\over R^2}\nabla\cdot(R^2\nabla\S_0), 
\l{identity}\ee 
which holds for any constant $\alpha$ and any functions $R$ and $\S_0$. Then, 
if $R$ satisfies the continuity equation 
\be 
\nabla\cdot(R^2\nabla\S_0)=0, 
\l{conteq}\ee 
and setting $\alpha=i/\hbar$, we have 
\be 
{1\over2m}(\nabla\S_0)^2=-{\hbar^2\over2m}{\Delta(Re^{{i\over 
\hbar}\S_0})\over Re^{{i\over\hbar}\S_0}}+{\hbar^2\over2m}{\Delta R\over R}. 
\l{identity2}\ee 
Comparing Eq.(\ref{aa10bbbbb}) and (\ref{oleole}) with Eq.(\ref{identity2}) we 
can make the following identification 
\be 
\W(q)=V(q)-E={\hbar^2\over2m}{\Delta(Re^{{i\over\hbar}\S_0})\over Re^{{i\over 
\hbar}\S_0}}, 
\l{evvaicormambo}\ee 
\be 
Q(q)=-{\hbar^2\over2m}{\Delta R\over R}. 
\l{identity3}\ee 
 
Let us now consider an arbitrary state, not necessarily corresponding to a 
$\W$ of the kind (\ref{ftpo}), with some reduced action $\S_0$. We consider 
$R$ solution of (\ref{conteq}). Note that, as (\ref{identity}) is independent 
of the form of $\W$, we have that (\ref{identity2}) holds for arbitrary $\S_0$ 
and $R$ satisfying (\ref{conteq}). We now start showing that (\ref{identity3}) 
holds in general, not only in the case (\ref{ftpo}). 
 
Let us set 
\be 
\W(q)={\hbar^2\over2m}{\Delta(Re^{{i\over\hbar}\S_0})\over Re^{{i\over\hbar}\S_0
}}+g(q). 
\l{daielo}\ee 
Eqs.(\ref{aa10bbbb})(\ref{identity2}) and (\ref{daielo}) imply 
\be 
Q(q)=-{\hbar^2\over2m}{\Delta R\over R}-g(q). 
\l{daielo2}\ee 
 
We have seen in (\ref{rota}) that the system described by ${\tilde\S_0}(\tilde 
q)$, where $\tilde q=\Lambda q+b$, has the important property that $\tilde\W( 
\tilde q)=\W(q(\tilde q))$ and $\tilde Q(\tilde q)=Q(q(\tilde q))$. Furthermore, 
using $\tilde\S_0(\tilde q)=\S_0(q)$, we find 
\be 
\tilde\nabla \cdot ({\tilde R}^2(\tilde q)\tilde\nabla{\tilde\S}_0(\tilde 
q))=\nabla \cdot ({\tilde R}^2(\tilde q(q)) \nabla\S_0(q))=0. 
\l{bo2}\ee 
Now observe that the continuity equation implies that ${\tilde R}^2(\tilde q(q) 
)\nabla\S_0(q)$ is the curl of some vector. In general we have $R^2 
\partial_i\S_0=\epsilon_i^{\;\,i_2\ldots i_D}\partial_{i_2}F_{i_3\ldots i_D}$, 
where $F$ is a $(D-2)$--form. Later we will exploit the field $F$. Therefore 
${\tilde R}^2(\tilde q(q))\nabla\S_0(q)$ must be a vector. On the other hand, 
since also 
$\nabla\S_0$ is a vector, we have that 
$\tilde R(\tilde q(q))$ must be a scalar under rotations and 
translations 
\be 
\tilde R(\tilde q)=R(q(\tilde q)). 
\l{bo4}\ee 
in agreement with the fact that $\nabla \cdot (R^2(q)\nabla\S_0(q))=0$. 
Therefore, we have 
\be 
\tilde g(\tilde q)=-{\hbar^2\over2m}{\tilde\Delta\tilde R\over\tilde R}- 
\tilde Q(\tilde q)=-{\hbar^2\over2m}{\Delta R\over R}-Q(q)=g(q), 
\l{bo6}\ee 
that is $g$ is scalar under rotations and translations. This implies that $g$ 
may depend only on $(\nabla\S_0)^2$, $\Delta\S_0$, $R$, $\Delta R$, $(\nabla 
R)^2$ and higher  derivatives which are invariant under rotations and 
translations, that is\footnote{A  possible dependence of $g$ on  
$\S_0$ would imply, against Eq.(\ref{traslazione}), that $\S_0$ satisfies a
differential equation involving $\S_0$ itself.} 
\be 
g=H((\nabla\S_0)^2,\Delta\S_0,R,\ldots). 
\l{owijfcw}\ee 
Let us now consider the case in which $\W=\sum_{k=1}^D\W_k(q_k)$, so that the 
problem reduces to a one dimensional one. In this case we have $g=0$, 
$\S_0=\sum_{k=1}^D\S_{0,k}(q_k)$ and $R=\prod_{k=1}^DR_k(q_k)$, 
where $R_k\propto(\partial_q\S_{0,k})^{-1/2}$. This provides the following 
constraints
\be 
H\left(\sum_k(\partial_k\S_{0,k})^2,\sum_k\partial_k^2\S_{0,k}, 
\prod_{k}R_k(q_k),\ldots\right)=0. 
\l{owijfcwaridaie}\ee 
This implies that $H\equiv0$. Let us show this in two related way. First note 
that by summing the one dimensional QSHJE and then performing a rotation, we 
arrive to a 
$D$--dimensional QSHJE which does not decompose in 1D QSHJE in the transformed 
coordinates. Thus (\ref{owijfcwaridaie}) also implies  
$H((\nabla\S_0)^2,\Delta\S_0,R,\ldots)=0$. 
 
A similar reasoning to prove that $H$ vanishes identically is to note that by  
(\ref{owijfcw}) and (\ref{owijfcwaridaie}) the only possibility to have a 
non--trivial $H$ is that it depends in a suitable way on terms that cancel 
when $\S_0=\sum_{k=1}^D\S_{0,k}(q_k)$. The building blocks to construct such 
terms have the form $\partial_j\partial_k\S_0$, $j\ne k$, which vanish when 
$\S_0=\sum_{k=1}^D\S_{0,k}(q_k)$. On the other hand, 
such terms are not scalar under rotations, so that they may enter in $H$ only 
if suitably saturated with other indices. Since the only vectorial indices at 
our disposal  are provided by derivatives, we see that there are not terms 
which are scalar under  rotations and vanish identically when 
$\W=\sum_{k=1}^D\W_k(q_k)$. Hence 
\be 
g=0. 
\l{abbastanzanullo}\ee 
 
Therefore, we have the basic result that the EP actually implies that in any 
dimension the reduced action satisfies the QSHJE 
\be 
{1\over2m}(\nabla\S_0)^2+\W-{\hbar^2\over2m}{\Delta R\over R}=0, 
\l{QSHJEHD}\ee 
and the continuity equation 
\be 
\nabla \cdot(R^2\nabla\S_0)=0. 
\l{conteqHD}\ee 
This equation implies the $D$--dimensional Schr\"odinger equation 
\be 
\left[-{\hbar^2\over2m}{\Delta}+V(q)\right]\psi=E\psi. 
\l{yz1xxxx5}\ee 
We stress that also in the higher dimensional case there is a fundamental 
difference between the correspondence (\ref{QSHJEHD})(\ref{conteqHD}) and 
(\ref{yz1xxxx5}) and the one usually considered in the literature. Namely, we 
have seen in the one--dimensional case that in general 
$Re^{{i\over\hbar}\S_0}$ cannot be identified with the wave--function. In 
particular, this would cause trouble in the case of bound states, as 
$\S_0$ would be a constant and inconsistencies arise in the classical limit.
This was also evident from the fact that $\{\S_0,q\}$ is not defined for 
$\S_0=cnst$. In the higher dimensional case, this would lead to the 
degeneration of the continuity equation (\ref{conteqHD}), with (\ref{QSHJEHD}) 
resulting in 
\be 
\W={\hbar^2\over2m}{\Delta R\over R}, 
\l{QSHJEHD2}\ee 
that in the classical limit, that by definition corresponds to $Q=0$, would 
lead to the contradiction 
\be 
\W=0. 
\l{contradiction}\ee 
Therefore, we have 
 
\vspace{.333cm} 
 
\noindent 
{\it The general relationship between the wave--function, $R$ and $\S_0$ has 
the form}
\be 
\psi=R\left(A e^{-{i\over\hbar}\S_0}+Be^{{i\over\hbar}\S_0}\right). 
\l{unafeaturadistinguente}\ee 
{\it In particular, for bound states we have} 
\be 
|A|=|B|. 
\l{amodulougualeab}\ee 
{\it Furthermore, $\S_0$ is never a constant.} 
 
\vspace{.333cm} 
 
Finally, we note that by Eqs.(\ref{azzo2yyyxxaa10bbbb}) and (\ref{identity3}) 
\be 
\Z(q^a;q^b)=(p^b|p^a)Q^a(q^a)-Q^b(q^b)=-{\hbar^2\over2m}\left[(p^b|p^a) 
{\Delta^aR^a\over R^a}-{\Delta^bR^b\over R^b}\right]. 
\l{cocicloide}\ee 
 
\subsection{Inversion II}\l{inversion2} 
 
$R^b(q^b)$ and $R^a(q^a)$ are related in a simple fashion in the case of 
rotations, reflections, dilatations and translations. Namely, $R^a(q^a)$ 
solves the continuity equation in the $q^a$--system if and only if 
$R^a(q^a(q^b))$ solves the continuity equation in the $q^b$--system. Thus, as 
we already proved in section 
\ref{mobius}, we explicitly verify by (\ref{cocicloide}) that $(q^a;q^b)$ 
vanishes identically in such cases. The analogous relation for the inversion is 
\be 
R^*(q^*)=r^{D-2}R(q)={1\over r_*^{D-2}}R(q(q^*)). 
\l{Rinversione}\ee 
Basically, we will show that, given a solution $R(q)$ of the continuity 
equation in the $q$--system, $R^*(q^*)$ in Eq.(\ref{Rinversione}) solves the 
continuity equation in the 
$q^*$--system. Then, we will verify that $(q;q^*)=0$, as we proved in section 
\ref{mobius} as a direct consequence of the EP. 
 
The continuity equation for $R^*(q^*)$ reads 
$$ 
\nabla^*\cdot({R^*}^2(q^*)\nabla^*\S^*_0(q^*))= 
$$ 
\be 
r^4\nabla\cdot({R^*}^2(q^*)\nabla\S_0(q))+(4-2D)r^2{R^*}^2\sum_{k=1}^Dq_k 
\partial_k\S_0(q)=0, 
\l{Rinvereq1}\ee 
that, after setting ${R^*}^2(q^*)=h(q)R^2(q)$, becomes 
\be 
hr^4\nabla \cdot \left(R^2\nabla\S_0\right)+ r^4R^2\nabla 
h\cdot\nabla\S_0+(4-2D)r^2R^2h\sum_{k=1}^Dq_k\partial_k\S_0=0, 
\l{zummoloa21}\ee 
implying by Eq.(\ref{bella2}) 
\be 
\left(r^2\nabla h+(4-2D)h\vec{q}\,\right)\cdot\nabla\S_0=r^{2D-2}\nabla(r^{4-2D} 
h)\cdot\nabla\S_0=0, 
\l{Rinvereq2}\ee 
which is solved by 
\be 
h(q)=r^{2D-4}. 
\l{Rinvereq3}\ee 
 
We now show that $(q=(q^*)^*;q^*)$ vanishes identically. By (\ref{cocicloide}) 
we have
\be 
(q;q^*)=(p_*|p)Q(q)-Q^*(q^*)=-{\hbar^2\over2m}\left(r^4{\Delta R\over R} 
-{\Delta^*R^*\over R^*}\right). 
\l{parentesi1}\ee 
On the other hand, by (\ref{Rinversione}) 
\be 
{\Delta^*R^*\over R^*}={\Delta(r^{D-2}R)\over r^{D-6}R}+(4-2D){\sum_{k=1}^D 
q_k\partial_k(r^{D-2}R)\over r^{D-4}R}=r^4{\Delta R\over R}, 
\l{bellagioia1}\ee 
so that 
\be 
(q;q^*)=0. 
\l{bellagioia3}\ee 
 
\mysection{Relativistic extension and Klein--Gordon equation}\l{questasezione} 
 
A basic property of the EP is that it has a universal character. In general, 
the implementation of the EP leads to a deformation of the corresponding 
classical HJ equation. In this respect, we note that existence of a fixed 
point in the non--relativistic stationary case demands the principle to be 
implemented in all the other circumstances. If we did not modify the 
time--dependent case as well, then taking the stationary limit would lead to 
inconsistencies. In other words, since modifying the stationary classical 
equation comes from a modification of the classical transformation properties 
of 
$\W$, which in general gets an inhomogeneous contribution, such as $(q;q^v)$, 
consistency implies that also in the time--dependent case the potential cannot 
transform as in the classical case. 
 
We start by deriving the Relativistic Quantum HJ Equation (RQHJE). Here we 
will consider the case in which the external potential is described by an 
arbitrary potential $V(q,t)$. This form will be particularly useful in 
deriving the time--dependent Quantum HJ Equation (QHJE), which in turn implies 
the time--dependent Schr\"odinger equation, as the non--relativistic limit of 
the RQHJE. Later on, we will consider the case in which the interaction is 
given in terms of the electromagnetic four--vector $A_\mu$. 
 
The Relativistic Classical Hamilton--Jacobi Equation (RCHJE) reads 
\be 
{1\over2m}\sum_{k=1}^D({\partial_k\S^{cl}(q,t)})^2+\W_{rel}(q,t)=0, 
\l{rel1}\ee 
where 
\be 
\W_{rel}(q,t)={1\over2mc^2}[\,m^2c^4-(V(q,t)+{\partial_t\S^{cl}(q,t)})^2]. 
\l{rel2}\ee 
In the time--independent case one has $\S^{cl}(q,t)=\S_0^{cl}(q)-Et$, and 
(\ref{rel1})(\ref{rel2}) become 
\be 
{1\over2m}\sum_{k=1}^D({\partial_k\S_0^{cl}})^2+\W_{rel}=0, 
\l{rel3}\ee 
and 
\be 
\W_{rel}(q)={1\over2mc^2}[\,m^2c^4-(V(q)-E)^2]. 
\l{rel4}\ee 
In the latter case, we can go through the same steps as in the 
non--relativistic case and the stationary RQHJE reads 
\be 
{1\over2m}(\nabla\S_0)^2+\W_{rel}-{\hbar^2\over2m}{\Delta R\over R}=0, 
\l{rel5}\ee 
where $R$ satisfies the continuity equation 
\be 
\nabla \cdot(R^2\nabla\S_0)=0. 
\l{rel5b}\ee 
Furthermore, (\ref{rel5})(\ref{rel5b}) imply the stationary Klein--Gordon 
equation
\be 
-{\hbar^2c^2}{\Delta\psi}+(m^2c^4-V^2+2EV-E^2)\psi=0, 
\l{rel6}\ee 
where $\psi=R\exp(i\S_0/\hbar)$. 
 
\subsection{Time--dependent case}\l{timecase} 
 
Let us start by noticing that in the time--dependent case, the 
$(D+1)$--dimensional RCHJE can 
be cast in the form (later on summation on repeated indices is understood)
\be 
{1\over2m}\eta^{\mu\nu}{\partial_\mu\S^{cl}}{\partial_\nu\S^{cl}}+
\W'_{rel}=0, 
\l{rel7}\ee 
where $\eta^{\mu\nu}$ is the Minkowski metric ${\rm diag}
\,(-1,1,\ldots,1)$, and 
\be 
\W'_{rel}(q)={1\over2mc^2}[\,m^2c^4-V^2(q)-2cV(q)\partial_0\S^{cl}(q)], 
\l{rel8}\ee 
where $q\equiv(ct,q_1,\ldots,q_D)$. We thus recognize that Eq.(\ref{rel7}) has
the same structure as Eq.(\ref{rel3}), the Euclidean metric being replaced by 
the Minkowskian one. Also in this case, in order to implement the EP, we have 
to modify the classical equation by adding a function to be determined, namely 
\be 
{1\over2m}(\partial\S)^2+\W_{rel}+Q=0. 
\l{rel7conlaQ}\ee 
Observe that since now $\W_{rel}'$ depends on $\S^{cl}$, we have to make the 
identification 
\be 
\W_{rel}(q)={1\over2mc^2}[\,m^2c^4-V^2(q)-2cV(q)\partial_0\S(q)], 
\l{rel8dues}\ee 
which differs from $\W_{rel}'$ for the Hamiltonian principal function as now 
$\S$ appears rather than $\S^{cl}$. 
 
Implementation of the EP requires that for an arbitrary $\W^a$ state 
\be 
\W^b_{rel}(q^b)=(p^b|p^a)\W^a_{rel}(q^a)+\Z(q^a;q^b), 
\l{rel9}\ee 
and 
\be 
Q^b(q^b)=(p^b|p^a)Q^a(q^a)-\Z(q^a;q^b), 
\l{rel10}\ee 
where this time 
\be 
(p^b|p)={\eta^{\mu\nu}p_\mu^bp_\nu^b\over\eta^{\mu\nu}p_\mu p_\nu}= {p^tJ\eta
J^tp\over p^t\eta p}, 
\l{rel11}\ee 
and $J$ is the Jacobian matrix 
\be 
{J^\mu}_\nu={\partial q^\mu\over\partial{q^b}^\nu}. 
\l{lajacobiannaaae}\ee 
Furthermore, we obtain the cocycle condition 
\be 
(q^a;q^c)=(p^c|p^b)\left[(q^a;q^b)-(q^c;q^b)\right]. 
\l{cociclo4}\ee 
As reported in Appendix B, this cocycle condition and the condition
\be 
\F_{rel}(\partial_\mu\S,\Box\S,\ldots)=0,
\l{traslazionerelAPPENDIX}\ee
which is the  analogue of the condition (\ref{traslazione}), imply that
\be 
(\gamma(q);q)=0.
\l{eqqconformal}\ee

\subsection{The RQHJE} 
 
\noindent 
Let us now consider the following identity 
\be 
\alpha^2(\partial\S)^2={\Box(Re^{\alpha\S})\over Re^{\alpha\S}} 
-{\Box R\over R}-{\alpha\over R^2}\partial \cdot (R^2\partial\S), 
\l{identityrel}\ee 
which holds for any constant $\alpha$ and any functions $R,\S$. Then, if
$R$ satisfies the continuity equation $\partial(R^2\cdot\partial\S)=0$, and 
setting $\alpha=i/\hbar$ we have 
\be 
{1\over2m}(\partial\S)^2=-{\hbar^2\over2m}{\Box(Re^{{i\over 
\hbar}\S})\over Re^{{i\over\hbar}\S}}+{\hbar^2\over2m}{\Box R\over R}. 
\l{identity2rel}\ee 
Our aim is to prove that 
\be 
\W_{rel}={\hbar^2\over2m}{\Box(Re^{{i\over\hbar}\S})\over
Re^{{i\over\hbar}\S}},
\l{ilwprimorel}\ee 
that by (\ref{identity2rel}) implies 
\be 
Q_{rel}=-{\hbar^2\over2m}{\Box R\over R}. 
\l{identity3rel}\ee 
Suppose that 
\be 
\W_{rel}={\hbar^2\over2m}{\Box(Re^{{i\over\hbar}\S})\over
Re^{{i\over\hbar}\S}}
+g, 
\l{daielorel}\ee 
and correspondingly 
\be 
Q_{rel}=-{\hbar^2\over2m}{\Box R\over R}-g. 
\l{daielo2rel}\ee 
First of all, we have seen that the system described by ${\tilde\S} (\tilde 
q)=\S(q)$, where $\tilde q=\Lambda q+b$, has the important property that 
$\tilde 
\W_{rel}(\tilde q)=\W_{rel}(q)$. Furthermore, we find 
\be 
\tilde\partial \cdot ({\tilde R}^2\tilde\partial\tilde\S)=\partial\cdot( 
{\tilde R}^2\partial\S)=0. 
\l{bo2rel}\ee 
Comparing this with the continuity equation
$\partial\cdot(R^2\partial\S)=0$, we have that under Poincar\'e
transformations, the $R$--function of the 
transformed system defined by $\tilde\S(\tilde q)=\S(q)$ has the 
transformation property 
\be 
\tilde R(\tilde q)=R(q). 
\l{bo4rel}\ee 
Therefore, by (\ref{eqqconformal}) and (\ref{bo4rel}) we have 
\be 
\tilde g=-{\hbar^2\over2m}{\tilde\Box\tilde R\over\tilde R}-\tilde 
Q=-{\hbar^2\over2m}{\Box R\over R}-Q=g, 
\l{bo6rel}\ee 
that is $g$ is a scalar under the Poincar\'e transformations. This implies 
that $g(q)$ may depend only on $\Box{\S_0}$, $(\partial\S_0)^2$, $R$,
$\Box R$ and $(\partial R)^2$ and higher derivatives which are
invariant under Poincar\'e transformations. However, in the time--independent 
limit the RQHJE must reduce to (\ref{rel5}), in particular 
$Q_{rel}\longrightarrow Q$. Therefore, $g$ must vanish in this limit implying 
\be 
g=0. 
\ee 
Then, the RQHJE reads 
\be 
{1\over2m}(\partial\S)^2+\W_{rel}-{\hbar^2\over2m}{\Box R\over R}=0, 
\l{bo7rel}\ee 
where $R$ and $\S$ satisfy the continuity equation 
\be 
\partial\cdot(R^2\partial\S)=0. 
\l{conteqrel}\ee 
 
\subsection{Non--relativistic limit} 
 
In this section we will consider the non--relativistic limit of the RQHJE. 
This will yield the time--dependent non--relativistic QHJE together with the 
time--dependent Schr\"odinger equation. 
 
To perform the classical limit we first need to make the usual substitution 
$\S=\S'-mc^2t$ and then taking the limit $c\longrightarrow\infty$. We have 
\be 
\W_{rel}\longrightarrow{1\over2}mc^2+V, 
\l{bosellane}\ee 
\be 
-{1\over2m}({\partial_0}\S)^2\longrightarrow{\partial\over\partial t}\S'- 
{1\over2}mc^2,
\ee 
\be 
\partial\cdot(R^2\partial\S)=0\longrightarrow m{\partial\over\partial 
t}R'^2+\nabla\cdot(R'^2\nabla\S')=0. 
\l{yeah}\ee 
Therefore, in the non--relativistic limit Eq.(\ref{bo7rel}) becomes (we remove
the $'$ from $R$ and $\S$) 
\be 
{1\over2m}(\nabla\S)^2+V+{\partial\over\partial t}\S-{\hbar^2\over2m}{\Delta
R\over R}=0,
\l{lanonrelllaaa}\ee
with the time--dependent non--relativistic continuity equation being 
\be 
m{\partial\over\partial t}R^2+\nabla\cdot(R^2\nabla\S)=0. 
\l{dajie}\ee 
It is then easy to see that (\ref{lanonrelllaaa}) and (\ref{dajie}) imply 
\be 
i\hbar{\partial\over\partial t}\psi=\left(-{\hbar^2\over2m}\Delta+V\right)\psi, 
\l{yeah2}\ee 
where $\psi=R\exp(i\S/\hbar)$. Note that if we used $\psi=R\exp(-i\S/\hbar)$, 
then we would get the complex conjugate of (\ref{yeah2}). 
 
\mysection{Gauge invariance and EP}\l{gauge} 
 
In section \ref{questasezione}, we derived the RQHJE with an arbitrary 
potential. As a byproduct, we obtained the time--dependent Schr\"odinger 
equation in the non--relativistic limit. This was a nice step as it indicates 
that, to be correctly implemented, the EP must be formulated in the exact 
framework, that is the relativistic one. In other words, even if the 
time--dependent Schr\"odinger equation can be derived directly in the 
non--relativistic framework \cite{6}, the natural realm to implement the EP is 
not in the approximate theory. In fact, while the Klein--Gordon equation 
follows naturally from the EP, the derivation of the time--dependent 
Schr\"odinger equation is less straightforward if one derives it directly in 
the non--relativistic theory. 
 
However, even if the derivation of the RQHJE is perfectly consistent, the 
formulation becomes particularly transparent if one works with gauge theories. 
 
\subsection{Minimal coupling from the EP} 
 
The point is that in general we considered $\W$ as an external fixed quantity,
then the corrections concerned $\S^{cl}$, as $\S$ solves an equation which is 
modified by the quantum potential. Nevertheless, we saw that if one considers 
the relativistic extension, then $\W$ contains 
$\S$ itself. So special relativity leads to consider $\W$ as composed 
by an external potential and $\S$ (see (\ref{rel8})). On the other hand, 
standard QM problems generally correspond to effective potentials. So, for 
example, the potential well, does not exist as a fundamental interaction. 
Thus, the nature of the EP indicates that it should be formulated in the 
framework of fundamental interactions. On the other hand, since we are in the 
relativistic framework, interactions cannot be strictly separated in kinetic 
and potential part. So the only possibility is that both are included in a 
generalized kinetic term, with $\W$ being space--time independent. It is clear
that this fixes the interaction to be described in terms of the minimal 
coupling. On the other hand, the minimal coupling prescription is at the heart
of gauge theories. 
 
We now show how the EP is simply implemented once one considers the minimal 
coupling prescription. Let us consider the interaction to be described in 
terms of the electromagnetic four--vector $A_\mu$. Let us set 
$P_\mu^{cl}=p_\mu^{cl}+eA_\mu$ where $p_\mu^{cl}$ is particle's momentum and
$P_\mu^{cl}=\partial_\mu\S^{cl}$ is the generalized one. In this case the
RCHJE reads
\be 
{1\over2m}(\partial\S^{cl}-eA)^2+{1\over2}mc^2=0, 
\l{ola1}\ee 
where $A_0=-{V\over ec}$. Note that now 
\be 
\W={1\over2}mc^2, 
\l{wugualemc2}\ee 
and the critical case corresponds to the limit situation in which $m=0$. As 
usual, in order to implement the EP, we are forced to add a correction to 
(\ref{ola1}) 
\be 
{1\over2m}(\partial\S-eA)^2+{1\over2}mc^2+Q=0. 
\l{ola1xbisse}\ee 
Furthermore, we have the transformation properties
\be
\W^b(q^b)=(p^b|p^a)\W^a(q^a)+\Z(q^a;q^b), 
\l{rel9A}\ee 
and 
\be
Q^b(q^b)=(p^b|p^a)Q^a(q^a)-\Z(q^a;q^b), 
\l{rel10A}\ee 
where 
\be 
(p^b|p)={(p^b-eA^b)^2\over(p-eA)^2}={(p-eA)^tJ\eta J^t(p-eA) 
\over(p-eA)^t\eta(p-eA)}, 
\l{rel11A}\ee 
and $J$ is the Jacobian matrix 
\be 
{J^\mu}_\nu={\partial q^\mu\over\partial{q^b}^\nu}. 
\l{opsiqj}\ee 
These transformations imply the cocycle condition 
\be 
(q^a;q^c)=(p^c|p^b)\left[(q^a;q^b)-(q^c;q^b)\right]. 
\l{cociclo4A}\ee 
As we proved in subsection \ref{timecase}, $(q^a;q^b)$ vanishes if $q^a$ and
$q^b$ are related by a conformal transformation. 
 
As usual we now have to consider the relevant identity for the (generalized) 
kinetic term. We have 
\be 
\alpha^2(\partial\S-eA)^2={D^2Re^{\alpha\S}\over Re^{\alpha\S}} 
-{\Box R\over R}-{\alpha\over R^2}\partial\cdot(R^2(\partial S-eA)). 
\l{insertinuccettino}\ee 
where 
\be 
D_{\mu}=\partial_{\mu}-\alpha eA_{\mu}, 
\l{ilcovariante}\ee 
and 
\be 
D^2\equiv D^{\mu}D_{\mu}=\Box-2\alpha eA\partial+\alpha^2e^2A^2- 
\alpha e(\partial A). 
\l{ilquadrato}\ee 
Since the identity (\ref{insertinuccettino}) holds for any $R$, $\S$ and 
$\alpha$, we can require $\partial\cdot(R^2(\partial\S-eA))=0$, and then set 
$\alpha=i/\hbar$ to have 
\be 
(\partial\S-eA)^2=\hbar^2\left({\Box R\over R}-
{D^2(Re^{{i\over\hbar}\S})\over Re^{{i\over\hbar}\S}}\right). 
\l{conalphaugualeadiaribisse}\ee 
We stress that there is no loss of generality in considering 
(\ref{conalphaugualeadiaribisse}) since, by 
$\partial\cdot(R^2(\partial\S-eA))=0$, this is an identity. 
 
We now show that 
\be 
\W={\hbar^2\over2m}{D^2(Re^{{i\over\hbar}\S})\over Re^{{i\over\hbar}\S}}, 
\l{precedequellasotto}\ee 
which, by (\ref{ola1xbisse}) and (\ref{conalphaugualeadiaribisse}) implies 
\be 
Q=-{\hbar^2\over2m}{\Box R\over R}. 
\l{identity3relA}\ee 
To prove (\ref{precedequellasotto}) we first define the function $g(q)$ by 
\be 
\W={\hbar^2\over2m}{D^2(Re^{{i\over\hbar}\S})\over Re^{{i\over\hbar}\S}}+g. 
\l{daielorelA}\ee 
Similarly to the case of the function $g$ in Eq.(\ref{bo6rel}), also here $g$ 
is a scalar function under Poincar\'e transformations. Furthermore, since in 
the 
$A\longrightarrow0$ limit we should reproduce Eq.(\ref{bo7rel}) with 
$\W_{rel}=mc^2/2$, we see that 
\be 
g=0, 
\ee 
and the RQHJE reads 
\be 
(\partial\S-eA)^2+m^2c^2-\hbar^2{\Box R\over R}=0, 
\l{bo7relA}\ee 
where $R$ and $\S$ satisfy the continuity equation 
\be 
\partial\cdot(R^2(\partial\S-eA))=0. 
\l{conteqrelA}\ee 
Let us stress that the same result can be directly obtained from 
(\ref{rel1})(\ref{rel2}) by observing that (\ref{rel7}) coincides with 
(\ref{ola1}) after setting $\W_{rel}=mc^2/2$ and replacing 
$\partial_\mu\S^{cl}$ by 
$\partial_\mu\S^{cl}-eA_\mu$. 
 
One can check that Eqs.(\ref{bo7relA})(\ref{conteqrelA}) imply the 
Klein--Gordon equation
\be 
(i\hbar\partial+eA)^2\psi+m^2c^2\psi=0, 
\l{bo9relA}\ee 
where 
\be 
\psi=Re^{{i\over\hbar}\S}. 
\l{lapsiiii}\ee 
If we considered $\psi=Re^{-{i\over\hbar}\S}$, then we would have the complex 
conjugate of (\ref{bo9relA}) 
\be 
(i\hbar\partial-eA)^2\psi+m^2c^2\psi=0. 
\l{bo10relA}\ee 
 
In the time--independent limit $A_\mu=(-{V\over ec},0,\ldots,0)$, $\partial_t 
V=0$, both (\ref{bo9relA}) and (\ref{bo10relA}) reduce to the stationary 
Klein--Gordon equation (\ref{rel6}). Correspondingly, Eq.(\ref{conteqrelA}) 
reduces to the stationary continuity equation (\ref{rel5b}) 
\be 
\partial\cdot(R^2(\partial\S-eA))=\nabla\cdot(R^2\nabla\S_0)-{1\over c^2} 
(\partial_t R^2(V-E)+R^2\partial_tV)=\nabla\cdot(R^2\nabla\S_0)=0. 
\l{rel7A}\ee 
 
\subsection{EP and mass generation} 
 
A special property of the EP is that it cannot be implemented in CM because of 
the fixed point corresponding to $\W^0\equiv0$. Implementing the EP then 
forces us to introduce a univocally determined piece to the classical HJ 
equation. A remarkable fact is that in the case of the RCHJE (\ref{ola1}), the 
fixed point 
$\W^0(q^0)\equiv0$ corresponds to $m=0$. The EP then implies that from this all 
the other masses can be generated by a coordinate transformation. Thus, we have 
 
\vspace{.333cm} 
 
\noindent 
{\it Masses correspond to the inhomogeneous term in the transformation 
properties of the $\W^0$ state} 
\be 
{1\over2}mc^2=\Z(q^0;q). 
\l{lamassa}\ee 
{\it Furthermore, by (\ref{rel9A}) (\ref{rel10A}) masses are expressed in 
terms of the quantum potential} 
\be 
{1\over2}mc^2=(p|p^0)Q^0(q^0)-Q(q). 
\l{lamassadaiquanti}\ee 
 
\vspace{.333cm} 
 
A basic feature of the formulation is that the EP implies that $\S$ is never 
trivial. So, for example also in the case of the non--relativistic particle 
with $V-E=0$, we have a non--trivial quantum potential 
\cite{1}--\cite{6}. In particular, in \cite{6} the role of the quantum 
potential was seen as a sort of intrinsic self--energy which is reminiscent of 
the relativistic self--energy. Eq.(\ref{lamassadaiquanti}) provides a more 
explicit evidence of such an interpretation. 
 
Furthermore, in \cite{5,6} it has been shown that tunnelling is a direct 
consequence of the quantum potential. In particular, $Q$ provides the energy 
to make $p$ real, and can be seen as the response of the particle self--energy 
to external potentials. This example also shows that external potential and 
particle energy are strictly related, and so they should be considered as 
components of a single object. In part, this is what the minimal coupling 
prescription provides. However, the EP, which naturally leads to such a 
prescription, also implies the additional quantum potential. In 
\cite{1}--\cite{6} it has been shown that this contribution is not fixed, 
rather it may change once the ``hidden variables'' are changed. In particular, 
a change of $\ell$ corresponds to a mixing between the $p^2/2m$ term and the 
quantum potential. We now show that in higher dimension there is a new degree 
of freedom, represented by an antisymmetric tensor, which is related to the 
hidden variables. 
 
\subsection{EP and the hidden antisymmetric tensor of QM} 
 
A basic property of the formulation immediately appears in one dimension once 
we consider the QSHJE. To understand this point it is useful to recall that 
the difference between the QSHJE and the one considered by Bohm, is that the 
QSHJE is written in terms of one function only: the reduced action $\S_0$. 
While we always have $\S_0\ne cnst$, in Bohm theory one has $\S_0=cnst$ for 
bound states. However, if one excludes, as implied by the EP, the trivial 
solution, then one can obtain the QSHJE from the standard version. In doing 
this one has to express $R$ in terms of $\S_0$ by solving the continuity 
equation. We now show that if one tries to write down the QHJE in higher 
dimension by solving the continuity equation, then a new field appears. We 
already encountered this situation in subsection 
\ref{uccellagione}. Namely, we saw that the continuity equation of the QSHJE 
implies that $R^2\nabla\S_0$ is given by the generalized curl of a 
$(D-2)$--form $F$. More precisely, we saw that 
\be 
R^2\partial_i\S_0=\epsilon_i^{\;\,i_2\ldots i_D}\partial_{i_2}F_{i_3\ldots 
i_D}.
\l{EPGAUGE1}\ee 
This equation is equivalent to 
\be 
R^2\partial_i\S_0=\partial^jB_{ij}, 
\l{EPGAUGE2}\ee 
where $B_{ij}$ is the antisymmetric two--tensor 
\be 
B_{ij}=\epsilon_{ij}^{\;\;\,\,i_3\ldots i_D}F_{i_3\ldots i_D}. 
\l{EPGAUGE3}\ee 
In other words, the 2--form $B$ is the Hodge dual of $F$ 
\be 
B=*F, 
\l{EPGAUGE4}\ee 
and the continuity equation is 
\be 
{\rm d}^+{\rm d}^+*F=*{\rm d}{\rm d}F=0. 
\l{EPGAUGE5}\ee 
In the time--dependent relativistic case $F$ is a $(D-1)$--form. We have 
\be 
R^2(\partial_\mu\S-eA_\mu)=\epsilon_\mu^{\;\,\sigma_1\ldots\sigma_D} 
\partial_{\sigma_1}F_{\sigma_2\ldots\sigma_D}=\partial^\nu B_{\mu\nu}, 
\l{EPGAUGE6}\ee 
that is 
\be 
R^2={(\partial^\mu\S-eA^\mu)\over(\partial\S-eA)^2}\partial^\nu B_{\mu\nu}, 
\l{EPGAUGE7}\ee 
or, equivalently, 
\be 
R^4={\partial^\nu B_{\mu\nu}\partial_\sigma B^{\mu\sigma}\over(\partial 
\S-eA)^2}. 
\l{EPGAUGE7b}\ee 
In terms of $B$ and $R$ the RQHJE (\ref{bo7relA}) reads 
\be 
\partial^\nu B_{\mu\nu}\partial_\sigma B^{\mu\sigma}+R^4m^2c^2 
-\hbar^2R^3\Box R=0. 
\l{EPGAUGE8}\ee 
 
The EP itself and the appearance of a new field indicates that now the RQHJE 
should be considered in a different context with respect to the usual one. So, 
for example, one may wonder if the $B$--field may help in considering a 
possible quantum origin of fundamental interactions. In the Introduction we 
suggested that QM and GR are facets of the same medal. More generally one 
should understand if there is a possible role of QM underlying the fundamental 
interactions or, more precisely, whether the EP underlies the structure of 
fundamental interactions trhough QM. We already saw evidence of the dynamical 
role of QM through the quantum potential \cite{6}, 
$e.g.$ in considering the tunnel effect. In this context one 
should understand whether Eq.(\ref{EPGAUGE8}) may provide a different 
understanding of the usual problems one meets in considering the Klein--Gordon 
equation. 
 
Let us find how $R$ transforms under general $v$--maps, $q\longrightarrow\tilde 
q(q)=v(q)$. In the $\tilde q$ system we have 
\be 
\tilde F_{\sigma_2\ldots\sigma_D}={\partial q^{\nu_2}\over\partial\tilde q^{ 
\sigma_2}}\ldots{\partial q^{\nu_D}\over\partial\tilde q^{\sigma_D}}F_{ 
\nu_2\ldots\nu_D}, 
\ee 
therefore 
$$ 
\epsilon^{\sigma_0\ldots\sigma_D}(\tilde\partial_{\sigma_0}\tilde\S-
e\tilde A_{\sigma_0})\tilde\partial_{\sigma_1}\tilde 
F_{\sigma_2\ldots\sigma_D}=\epsilon^{ 
\sigma_0\ldots\sigma_D}{\partial q^{\nu_0}\over\partial\tilde q^{\sigma_0}}(
\partial_{\nu_0}\S-eA_{\nu_0}){\partial q^{\nu_1}\over\partial\tilde q^{ 
\sigma_1}}{\partial q^{\nu_2}\over\partial\tilde q^{\sigma_2}}\ldots{\partial 
q^{\nu_D}\over\partial\tilde q^{\sigma_D}}\partial_{\nu_1}
F_{\nu_2\ldots\nu_D}=
$$ 
\be 
\det\left({\partial q\over\partial\tilde q}\right)\epsilon^{\nu_0\ldots\nu_D} 
(\partial_{\nu_0}\S-eA_{\nu_0})\partial_{\nu_1}F_{\nu_2\ldots\nu_D}, 
\l{EPGAUGE10}\ee 
where we used the fact that the second derivatives of the Jacobian matrix get 
cancelled due to the antisymmetry of the Levi--Civita tensor. Finally, by 
(\ref{EPGAUGE6}) and (\ref{EPGAUGE7}), we have 
\be 
\tilde R^2=\det\left({\partial q\over\partial\tilde q}\right)(p|\tilde p)R^2, 
\l{EPGAUGE11}\ee 
that holds also in the stationary non--relativistic case with $(p|\tilde p)$ 
given by (\ref{natura2}). Given (\ref{EPGAUGE11}), we easily re--derive the 
transformation property of $R$ under the inversion $q^*(q)$ we derived in 
subsection \ref{inversion2}. In fact by (\ref{lappona1}) and (\ref{lappona2}) 
we have $\det^2(J^{-1})=\det(r^4{\bf 1}_D)$, that is 
\be 
{\rm det}^2\left({\partial q\over\partial q^*}\right)=\det(r^4{\bf 
1}_D)=r^{4D},
\l{EPGAUGE14}\ee 
then, as by (\ref{jacobinver}) $(p|p_*)=r^{-4}$, we obtain 
\be 
{R^*}^2(q^*)=r^{2D-4}R^2(q). 
\l{EPGAUGE13}\ee 
It is then convenient to use (\ref{EPGAUGE11}) to find the transformation 
property of $R$ under the inversion (\ref{rel13}). By (\ref{rel16b}) and 
(\ref{jacobinverRel}), we have 
\be 
{R^*}^2(q^*)=(q^2)^{D-1}R^2(q). 
\l{EPGAUGE15}\ee

\appendix

\mysection{}

In this appendix we report the proof of Eqs.(\ref{eqqq3})(\ref{eqq15b}) and 
(\ref{inver1e}) concerning the structure of the function
$(q^a;q^b)$ in the cases in which $q^a$ and $q^b$ are related by a
translations, dilatations and inversion respectively.

Let us start by considering the function 
\be 
G(D,q)=(q+D;q), 
\l{generale}\ee 
where $D$ is an arbitrary constant vector. In terms of $G(D,q)$, 
Eq.(\ref{eqq3}) yields 
\be 
G(D,q+B)-c(q+B+D)+c(q+D)=G(D,q)-c(q+B)+c(q). 
\l{eqq10fff}\ee 
Taking the derivative of both sides of Eq.(\ref{eqq10fff}) with respect to 
$B_j$, we get 
\be 
\partial_{q_j}G(D,q+B)-\partial_{q_j}c(q+B+D)=-\partial_{q_j}c(q+B). 
\l{eqq10f}\ee 
After setting $B_j=0$ and integrating 
\be 
G(D,q)=c(q+D)-c(q)+\hat{G}(D,\hat{q}), 
\l{eqq10g}\ee 
where, as before, by $\hat{q}$ we denote all the components of $q$ other than
$q_j$, and, by Eq.(\ref{eqq10e}), $\hat{G}(D,\hat{q})$ vanishes if all the 
components of $D$ other than $D_j$ are zero. Furthermore, plugging 
Eq.(\ref{eqq10g}) into Eq.(\ref{eqq2}), we realize that $\hat{G}(D,\hat{q})$ 
shares the same properties as 
$G(D,q)$. Therefore, the analogue of Eq.(\ref{eqq10g}) holds for
$\hat{G}(D,\hat{q})$ as well. Hence, applying this reasoning recursively we
end up with
\be 
(q+D;q)=F(q+D)-F(q)+H(D), 
\l{eqq10m}\ee 
where $H(D)$ vanishes whenever only one component of $D$ is not zero. However,
by (\ref{eqq2}), we find that $H$ is linear 
\be 
H(D+E)=H(D)+H(E), 
\l{eqq10i}\ee 
which implies 
\be 
H(D)=\sum_{k=1}^Da_kD_k, 
\l{eqq10j}\ee 
so that 
\be 
H=0, 
\l{hezero}\ee 
and we arrive to Eq.(\ref{eqqq3}), that is 
\be 
(q+D;q)=F(q+D)-F(q),
\l{eqq10hAPPENDIX}\ee 
Note that the right hand side remains invariant under the constant shift 
\be 
F\longrightarrow F+c. 
\l{paraponziperopero}\ee 
 
Let us now analyze the consequences of (\ref{eqq10hAPPENDIX}) on 
$h(A,q)=(Aq,q)$. By (\ref{cociclo3}) and noting that $(p|p_A)=A^2$, we have 
\be 
(A(q+B);q)=(A(q+B);q+B)+(q+B;q)=A^2 (Aq+AB;Aq)+(Aq;q), 
\l{eqq11}\ee 
which is equivalent to 
\be 
h(A,q+B)-h(A,q)=A^2[F(Aq+AB)-F(Aq)]-F(q+B)+F(q), 
\l{eqq12}\ee 
where now $B$ is an arbitrary vector. Taking the derivative with respect to 
$B_j$ 
\be 
\partial_{q_j}h(A,q+B)=A^3\partial_{q_j}F(Aq+AB)-\partial_{q_j}F(q+B), 
\l{eqq13}\ee 
and setting $B=0$ we find 
\be 
\partial_{q_j}h(A,q)=A^3\partial_{q_j}F(Aq)-\partial_{q_j}F(q), 
\l{eqq14}\ee 
that upon integration yields 
\be 
h(A,q)=A^2F(Aq)-F(q)+g(A). 
\l{eqq15}\ee 
A useful observation is that $h(A,q)=(Aq;q)$ evaluated at $q=0$ cannot depend
on $A$, so that $h(A,0)=h(1,0)=0$. Therefore 
\be 
g(A)=-(A^2-1)F(0), 
\l{eqq15a}\ee 
and $h(A,q)=A^2(F(Aq)-F(0))-(F(q)-F(0))$, that, upon the re--labeling
$F(q)\to F(q)-F(0)$, coincides with (\ref{eqq15b}), that is 
\be 
(Aq;q)=A^2F(Aq)-F(q), 
\l{eqq15bAPPENDIX}\ee 
where now 
\be 
F(0)=0. 
\l{FinzeroezeroAPPENDIX}\ee 
Note that this fixes the ambiguity (\ref{paraponziperopero}).

We now complete the proof of Eq.(\ref{inver1e}). First of all note that by 
(\ref{cociclo3}) 
\be 
((Aq)^*;q)=(p|p_A)((Aq)^*;Aq)+(Aq;q). 
\l{inver1a}\ee 
On the other hand, by (\ref{cociclo3}) and (\ref{dilatazioni}) 
\be 
((Aq)^*;q)=(A^{-1}q^*;q)=(p|p_*)(A^{-1}q^*;q^*)+(q^*;q), 
\l{inver1b}\ee 
so that 
\be 
A^2((Aq)^*;Aq)+(Aq;q)=r^{-4}(A^{-1}q^*;q^*)+(q^*;q), 
\l{insommadunque3}\ee 
and by (\ref{eqq15b}) 
\be 
A^2((Aq)^*;Aq)+A^2 F(Aq)-F(q)={1\over r^4}[A^{-2}F(A^{-1}q^*)-F(q^*)]+
(q^*;q).
\l{inver1c}\ee 
Picking a $q_0$ such that $q_0^*=q_0$ and noticing that $r_0=1$, we have by 
(\ref{eppercheno}) and (\ref{inver1c}) that 
\be 
((Aq_0)^*;Aq_0)=A^{-4}F(A^{-1}q_0)-F(Aq_0). 
\l{inver1d}\ee 
Now observe that any $q$ can be expressed as $Aq_0$, where $A=r$ and with 
$q_0$ a suitable solution of $q^*=q$. Furthermore, by (\ref{dilatazioni}) we 
have $A^{-1}q_0=(Aq_0)^*$, so that (\ref{inver1d}) is equivalent to 
Eq.(\ref{inver1e}), that is
\be 
(q^*;q)={1\over r^4}F(q^*)-F(q). 
\l{inver1eAPPENDIX}\ee

\mysection{}
 
In section \ref{mobius} we showed that, as a consequence of 
Eq.(\ref{cociclo3}), 
$(q^a;q^b)$ vanishes identically if $q^a$ and 
$q^b$ are related by a M\"obius transformation. In the present case, 
we can prove the analogous result that $(q^a;q^b)$ vanishes if $q^a$ and 
$q^b$ are related by a conformal transformation, where the conformal group, 
with respect to the Minkowski metric, is generated by translations 
\be 
q\longrightarrow q+b,\qquad b\in\RR^{D+1}, 
\l{Mrel1}\ee 
dilatations 
\be 
q\longrightarrow Aq,\qquad A\in\RR, 
\ee 
Lorentz transformations 
\be 
q\longrightarrow\Lambda q,\qquad\Lambda\in O(D,1), 
\l{rel12}\ee 
and the inversion 
\be 
q^*={q\over q^2}, 
\l{rel13}\ee 
where $q^2=\eta^{\mu\nu}q_\mu q_\nu$. Note that for the inversion to be 
well--defined, 
$\RR^{D+1}$ must be completed by a cone at infinity 
\cite{ItzZub}. This space is the analogue of $\hat\RR^{D+1}$. The proof
is the same as the one provided in section \ref{mobius} and in Appendix A. In 
particular, we have 
\be 
(q+B;q)=F(q+B)-F(q). 
\l{rel14}\ee 
\be 
(Aq;q)=A^2F(Aq)-F(q), 
\l{rel15}\ee 
\be 
(\Lambda q;q)=F(\Lambda q)-F(q), 
\l{rel16}\ee 
where $F$ is an arbitrary function satisfying $F(0)=0$. As far as the 
inversion is concerned, the proof needs to be slightly modified. The Jacobian
matrix of this mapping is given by 
\be 
{J^\mu}_\nu=\partial_\nu q^{*\mu}=\partial_\nu{q^\mu\over q^2}= 
{\delta^{\;\,\mu}_\nu\over q^2}-2{q^\mu q_\nu\over q^4}, 
\l{ermapping}\ee 
where $q^4\equiv(q^2)^2$. Then 
\be 
(J\eta J^t)^{\mu\nu}={J^\mu}_\rho\eta^{\rho\sigma}{J^\nu}_\sigma=\left({ 
\delta^{\;\,\mu}_\rho\over q^2}-2{q^\mu q_\rho\over q^4}\right)
\eta^{\rho\sigma}\left({\delta^{\;\,\nu}_\sigma\over q^2}-
2{q^\nu q_\sigma\over q^4}\right)= 
{1\over q^4}\eta^{\mu\nu}, 
\l{rel16b}\ee 
which implies 
\be 
(p|p_*)=(p_*|p)^{-1}={p_\mu\eta^{\mu\nu}p_\nu\over{p_*}_\mu\eta^{\mu\nu}{ 
p_*}_\nu}={p_*^tJ\eta J^tp_*\over p_*^t\eta p_*}={1\over q^4}. 
\l{jacobinverRel}\ee 
Note that $q^*$ is involutive since 
\be 
{q^*}^2={q^*}^\mu\eta_{\mu\nu}{q^*}^\nu={1\over q^4}q^\mu
\eta_{\mu\nu}q^\nu={1\over q^2}, 
\l{rdiefferel}\ee 
and therefore 
\be 
{(q^*)^*}^\mu={{q^*}^\mu\over{q^*}^2}={q^\mu\over q^2{q^*}^2}=q^\mu. 
\l{involutivarel}\ee 
Since $q^2$ is invariant under Lorentz transformations, these commute with the
inversion 
\be 
(\Lambda q)^*=\Lambda q^*. 
\l{trullallero}\ee 
Finally, under dilatations 
\be 
{(Aq)^*}^\mu={(Aq)^\mu\over q_A^2}={Aq^\mu\over A^2q^2}=A^{-1}{q^\mu\over q^2} 
=A^{-1}{q^*}^\mu, 
\l{dilatazionirel}\ee 
where $q_A^2=Aq^\mu\eta_{\mu\nu}Aq^\nu=A^2q^2$. By (\ref{jacobinverRel}) and 
(\ref{involutivarel}) 
\be 
(q^*;q)=-(p|p^*)(q;q^*)=-{1\over q^4}((q^*)^*;q^*), 
\l{iappalarel}\ee 
which implies that $(q^*;q)$ vanishes when evaluated at any $q_0$ solution of 
$q^*=q$ 
\be 
(q^*;q)|_{q=q_0}=0, 
\l{epperchenorel}\ee 
and that 
\be 
(q^*;q)|_{q=q_1}=-(q;q^*)|_{q=q_1}, 
\l{epperchenorel2}\ee 
where $q_1$ satisfies $q^*=-q$. By (\ref{cociclo4}) 
\be 
((Aq)^*;q)=(p|p_A)((Aq)^*;Aq)+(Aq;q). 
\l{inver1arel}\ee 
On the other hand, by (\ref{cociclo4}) and (\ref{dilatazionirel}) 
\be 
((Aq)^*;q)=(A^{-1}q^*;q)=(p|p_*)(A^{-1}q^*;q^*)+(q^*;q), 
\l{inver1brel}\ee 
therefore, by (\ref{rel15}) 
\be 
A^2((Aq)^*;Aq)+A^2F(Aq)-F(q)={1\over q^4}[A^{-2}F(A^{-1}q^*)-F(q^*)]+(q^*;q). 
\l{inver1crel}\ee 
Picking a $q_0$ such that $q_0^*=q_0$, Eq.(\ref{inver1crel}) yields 
\be 
((Aq_0)^*;Aq_0)=A^{-4}F(A^{-1}q_0)-F(Aq_0). 
\l{inver1drel}\ee 
Now observe that any $q$, such that $q^2>0$, can be expressed as $Aq_0$, where 
$A^2=q^2$, and with $q_0$ a suitable solution of $q^*=q$. Furthermore, by 
(\ref{dilatazionirel}) we have $A^{-1}q_0=(Aq_0)^*$, so that 
(\ref{inver1drel}) is equivalent to 
\be 
(q^*;q)={1\over q^4}F(q^*)-F(q),\qquad\qquad q^2>0. 
\l{inver1erel}\ee 
On the other hand, picking a $q_1$ such that $q_1^*=-q_1$, 
Eq.(\ref{inver1crel}) yields 
\be 
((Aq_1)^*;Aq_1)=A^{-4}F(A^{-1}q_1^*)-F(Aq_1)+A^{-2}(F(q_1)-F(q_1^*))+A^{-2} 
(q^*;q)|_{q=q_1}. 
\l{inver1drel2}\ee 
Taking $A=-1$, (\ref{inver1drel2}) becomes 
\be 
(q_1;q_1^*)=F(q_1)-F(-q_1)+F(q_1)-F(-q_1)+(q_1^*;q_1), 
\l{yellow}\ee 
which, by virtue of (\ref{epperchenorel2}), is equivalent to 
\be 
(q_1^*;q_1)=F(q_1^*)-F(q_1). 
\l{inver1erel2}\ee 
Hence, (\ref{inver1drel2}) becomes 
\be 
((Aq_1)^*;Aq_1)=A^{-4}F(A^{-1}q_1^*)-F(Aq_1). 
\l{inver1drel22}\ee 
Now observe that any $q$, such that $q^2<0$, can be expressed as $Aq_1$, where
$A^2=-q^2$, and with $q_1$ a suitable solution of $q^*=-q$. Furthermore, by 
(\ref{dilatazionirel}) we have $A^{-1}q_1^*=(Aq_1)^*$, so that 
(\ref{inver1drel22}) is equivalent to 
\be 
(q^*;q)={1\over q^4}F(q^*)-F(q),\quad q^2<0. 
\l{inver1erel22}\ee 
Thus, in general 
\be 
(q^*;q)={1\over q^4}F(q^*)-F(q). 
\l{inver1rel222}\ee 
However, also in the relativistic case we have the analogue of condition 
(\ref{traslazione}) 
\be 
\F_{rel}(\partial_\mu\S,\Box\S,\ldots)=0. 
\l{traslazionerel}\ee 
Then, the same reasoning as in subsection \ref{coboundary} leads to 
\be 
F=0. 
\l{acchiappa}\ee 
Therefore, we can state the following result 
 
\vspace{.333cm} 
 
\noindent 
{\it Eq.(\ref{traslazionerel}) and the cocycle condition (\ref{cociclo4}) 
imply that $(q^a;q^b)$ vanishes when $q^a$ and $q^b$ are related by a 
conformal transformation}
\be 
(\gamma(q);q)=0. 
\l{eqqconformalAPPENDIX}\ee 
 
\vspace{.333cm} 
 
\noindent 
{\bf Acknowledgements}. It is a pleasure to thank E.R. Floyd for several 
important discussions. We also thank M. Appleby, I. Bakas, D. Bellisai, M. 
Bochicchio, G. Bonelli, L. Bonora, R. Carroll, L. Cornalba, F. De Felice, G.F.
Dell'Antonio, E. Gozzi, F. Guerra, L.P. Horwitz, F. Illuminati, J.M. Isidro,
A. Kholodenko, A. Kitaev, P.A. Marchetti, G. Marmo, R. Nobili, R. Onofrio, F. 
Paccanoni, P. Pasti, P. Sergio, M. Tonin, G. Travaglini, G. Vilasi and R. 
Zucchini for stimulating discussions. 
 
G.B. is supported in part by DOE cooperative agreement DE--FC02--94ER40818 and
by INFN ``Bruno Rossi" Fellowship. G.B. also thanks the S.I.F. for the
``Alberto Frigerio" prize and the theory group of the University of Padova for
kind hospitality while working at the present paper. A.E.F. thanks the theory
divisions of CERN and ITP--UCSB and the theory group of the University of
Padova for kind hospitality while working at the present paper. A.E.F. is
supported in part by DOE Grant No.\ DE--FG--0287ER40328, and M.M. by the
European Commission TMR programme ERBFMRX--CT96--0045.

\end{document}